\documentclass[hyperlink,superscriptaddress,preprintnumbers,amsmath,amssymb,nofootinbib]{revtex4}
\usepackage{graphicx}
\usepackage{epstopdf}
\usepackage{dcolumn}
\usepackage{bm}
\usepackage{hyperref}
\usepackage{epsfig}
\def\slashchar#1{\setbox0=\hbox{$#1$} 
\dimen0=\wd0 
\setbox1=\hbox{/} \dimen1=\wd1 
\ifdim\dimen0>\dimen1 
\rlap{\hbox to \dimen0{\hfil/\hfil}} 
#1 
\else 
\rlap{\hbox to \dimen1{\hfil$#1$\hfil}} 
/ 
\fi}

\def\b{\beta}

\def\d{\delta}

\def\g{\gamma}

\def\l{\lambda}
\def\m{\mu}

\def\s{\sigma}

\def\G{\Gamma}

\def\L{\Lambda}

\def\beq{\begin{eqnarray}}
\def\eeq{\end{eqnarray}}

\newcommand{\vev}[1]{ \left\langle {#1} \right\rangle }

\def\tr{\mathop{\rm tr}}

\begin{document}
\title{Cosmic Ray Spectra in Nambu-Goldstone Dark Matter Models} 

\author{Masahiro Ibe}
\affiliation{%
SLAC National Accelerator Laboratory, Menlo Park, CA 94025
}%

\author{Hitoshi Murayama}
\affiliation{%
Department of Physics, University of California, Berkeley, 
and Theoretical Physics Group, LBNL, Berkeley, CA 94720 
}%
\affiliation{Institute for the Physics and Mathematics of the Universe, University of Tokyo, 
Kashiwa 277-8568, Japan.}
\author{Satoshi Shirai}
\affiliation{%
Department of Physics, University of Tokyo, Tokyo 113-0033, Japan.}
\affiliation{%
Institute for the Physics and Mathematics of the Universe, University of Tokyo, 
Kashiwa 277-8568, Japan.}%
\author{Tsutomu T.~Yanagida}
\affiliation{Institute for the Physics and Mathematics of the Universe, University of Tokyo, 
Kashiwa 277-8568, Japan.}
\affiliation{%
Department of Physics, University of Tokyo, Tokyo 113-0033, Japan.}

\begin{abstract}
We  discuss the cosmic ray spectra in 
annihilating/decaying Nambu-Goldstone dark matter models.
The recent observed positron/electron excesses at PAMELA and Fermi experiments 
are well fitted by the dark matter with a mass of 3\,TeV for the annihilating model, while 
with a mass of 6\,TeV for the decaying model.
We also show that the Nambu-Goldstone dark matter models predict
a distinctive gamma-ray spectrum in a certain parameter space.
\end{abstract}

\date{\today}
\maketitle
\preprint{SLAC-PUB-13688}
\section{Introduction}
Recent observations of the PAMELA\,\cite{Adriani:2008zr},
ATIC\,\cite{Chang:2008zz}, PPB-BETS\,\cite{Torii:2008xu},
and Fermi\,\cite{Abdo:2009zk} experiments strongly suggest the existence of a new source of  positron/electron fluxes.  
The most interesting candidate of the new source is
the dark matter  with a mass in the TeV range annihilating or decaying
into the visible particles which result in the positrons/electrons\,\cite{Bergstrom:2009ib}.
Especially, the Fermi experiment has released data on the electron/positron spectrum
from 20\,GeV up to 1\,TeV\,\cite{Abdo:2009zk}, where the spectrum falls as $E^{-3.0}$.
As reported in Refs.\,\cite{Murayama:2009, Bergstrom:2009fa,Meade:2009iu}
the data can be well fitted by the dark matter which mainly annihilates/decays into a pair of light scalars each of which subsequently decays into a pair of electrons or muons.%
\footnote{The earlier works on the cosmic ray spectra before the Fermi data
in the presence of the light scalars decaying into the light lepton pairs 
can be found in Ref.\,\cite{ArkaniHamed:2008qn,Nomura:2008ru}. 
}

From the theoretical point of view, it is always motivated to relate the identity of 
the dark matter with the 
new physics which is anticipated from other motivations\,\cite{Murayama:2007ek}.
Among them, one of the most motivated new physics is the supersymmetric standard model (SSM)
which is expected as a solution to the hierarchy problem.
In the SSM, the stable lightest supersymmetric particle (LSP) is a candidate 
of the dark matter.
The LSP interpretation with a mass in the TeV range\,\cite{Shirai:2009fq,Chen:2009mj}, 
however, 
implies that the masses of the other supersymmetric particles are much heavier than a TeV, 
which diminishes the significance of the SSM as a successful solution to the hierarchy problem.
Besides, such a heavy LSP interpretation can be falsified rather easily, once
the supersymmetric particles are discovered in hundreds GeV range
at the coming LHC experiments. 

The attempt to relate the dark matter to the supersymmetric models, however, 
should not be necessarily confined to the LSP dark matter scenarios.
In fact, the SSM always requires other new physics, the supersymmetry breaking sector
which may include a stable particle as a candidate of the dark matter.
The idea of the Nambu-Goldstone dark matter developed in Ref.\,\cite{Ibe:2009dx} is 
one of the realization of the dark matter in a supersymmetry breaking sector.
There, the dark matter is interpreted as a pseudo-Nambu-Goldstone boson
in a supersymmetry breaking sector so that the mass of the dark matter is in the TeV
range out of the dynamics of supersymmetry breaking at tens to hundreds TeV range.

In an explicit example of the Nambu-Goldstone dark matter scenario  given in 
Ref.\,\cite{Ibe:2009dx}, 
we considered the model where the dark matter annihilates into a pair of the light
pseudo scalars, the R-axions, via a narrow resonance, the flaton, which leads to
the right amount of the dark matter.
Furthermore, for the R-axion mainly decaying into a pair of electrons or muons, 
the final state of the dark matter is the four electrons or muons, which are favored
to explain the positron/electron spectrum observed by Fermi experiment.

The most prominent difference of the Nambu-Goldstone dark matter model
from the other dark matter models which explain the observed positron/electron excesses
is that the Nambu-Goldstone dark matter model is strongly interrelated to the physics of the SSM.
That is, in the Nambu-Goldstone dark matter scenario, 
it is difficult  for the supersymmetry breaking to be much higher than tens to hundreds TeV.
This restriction suggests that the supersymmetry breaking effects should be mediated 
to the SSM sector at the low energy scale, i.e., the model requires the gauge mediation 
mechanism\,\cite{Giudice:1998bp}. 
Thus, the dark matter interactions with the SSM particles  
are determined along with the gauge mediation effects.
With the interrelation to the SSM physics, the Nambu-Goldstone dark matter model has a distinctive prediction on such as a gamma-ray spectrum and an antiproton flux in cosmic ray. 

In this paper, we discuss the cosmic ray spectra 
in the Nambu-Goldstone dark matter scenario based
on the explicit model in Ref.\,\cite{Ibe:2009dx}.
As we will show, the Nambu-Goldstone dark matter model fits to the recently observed
positron/electron excesses in both the annihilating and the decaying dark matter scenarios.
Furthermore, we show that the model gives a distinctive prediction on a gamma-ray 
spectrum for a certain parameter space, which comes from 
the finite annihilation/decay rates of the dark matter into a pair of gluinos.

The organization of the paper is as follows.
 In section\,\ref{sec:model}, 
 we summarize the explicit example of the NDGM model in Ref.\,\cite{Ibe:2009dx}.
 There, we also derive conditions so that the dark matter density is consistent 
 with the observed dark matter density.
 In section\,\ref{sec:spectra}, we show how well the model fits the observed fluxes in both 
 the annihilating/decaying dark matter scenarios.
 We also demonstrate how the modes into the SSM particles affect the gamma-ray 
 spectrum.

\section{Nambu--Goldstone Dark Matter}\label{sec:model}
In this section, we review a model of the Nambu--Goldstone
dark matter model\,\cite{Ibe:2009dx} which is based on a vector-like SUSY breaking model in
Ref.\,\cite{Izawa:1996pk}.
In the model, the dark matters are interpreted as pseudo Nambu-Goldstone bosons
which result from spontaneous breaking of the approximate global
symmetry in the vector-like SUSY breaking model.

\subsection{Nambu-Goldstone Dark Matter, Flaton, and R-axion}
The key ingredients of the model are the light flaton which corresponds to the so-called 
pseudo moduli of the SUSY breaking model, and the R-axion
which is a pseudo Nambu-Goldstone boson resulting from spontaneous R-symmetry breaking.
In the followings, we overview the relevant properties of those  particles.

\subsubsection{SUSY breaking sector}
The vector-like SUSY breaking model is based on an $SU(2)$ gauge
theory with four fundamental representation fields
$Q_{i}(i=1,\cdots,4)$ and six singlet fields $S_{ij}=-S_{ji}$
($i,j=1,\cdots,4$).  In this model, the SUSY is
dynamically broken when the $Q$'s and $S$'s couple in the
superpotential,
\begin{eqnarray}
\label{eq:SQQ}
 W = \l_{ij} S_{ij}Q_{i}Q_{j}, \, (i<j),
\end{eqnarray}
where $\l_{ij}$ denote coupling constants.  
The maximal global symmetry this model may have is 
$SU(4)\simeq SO(6)$ symmetry which requires $\l_{ij} = \l$.
The SUSY is broken as a
result of the tension between the $F$-term conditions of $S$'s and
$Q$'s.  That is, the $F$-term conditions of $S_{ij}$, $\partial
W/\partial S_{ij} = \l_{ij}Q_{i}Q_{j} = 0$, contradict with the
quantum modified constraint ${\rm Pf}(M_{ij}) =\L_{\rm dyn}^{2}$ where
$M_{ij}$ denote composite gauge singlets made from $Q_i Q_j$.

Below the dynamical scale $\Lambda_{\rm dyn}$, the model is described by
the light degrees of freedom,  $M_{A}$ and $S_{A}$, ($A=1-6$), 
with a quantum modified constraint,
\begin{eqnarray}
\label{eq:M0}
\sum_{A=1-6} M_{A}^{2} = \L_{\rm dyn}^2.
\end{eqnarray}
Here, we have assumed that 
the effective composite operators $M_A$ are canonically normalized. 
Notice that the above quantum modified constraint  breaks
the global $SU(4)\simeq SO(6)$ symmetry into $SP(2)\simeq SO(5)$ symmetry.
Thus, when the model possesses the $SO(6)$ symmetry approximately,
the Nambu-Goldstone bosons appear as the result of the quantum  modified constraint.
Furthermore, the Nambu-Goldstone bosons have a long lifetime
when an appropriate subgroup of $SO(5)$ is almost exact.
(We will discuss more on the stability and the global symmetry in the next section.)
In this way, we realize a model with the Nambu-Goldstone dark matter with mass in
the  TeV range out of physics of tens to hundreds TeV range.
 
To make the discussion concrete, for a while, let us assume that the SUSY breaking sector possesses an $SO(5)\subset SO(6)$
global symmetry and rearrange the tree-level interaction Eq.\,(\ref{eq:SQQ}) so that the $SO(5)$
symmetry is manifest;
\begin{eqnarray}
\label{eq:SQQ2}
 W &=& \l\, S_{0}(QQ)_0 + \l' \sum_{a=1-5} S_{a}(QQ)_a\ ,\cr
 &=& \l\, \L_{\rm dyn}S_{0}M_0 + \l' \sum_{a=1-5} \L_{\rm dyn}S_{a}M_a\ .
\end{eqnarray}
In the second line, we rewrite the superpotential by using the low energy field $M_A$.%
\footnote{In the second line, we neglected order one coefficients of each term.}
We further assume $\l<\l'$ and assume that 
the model possesses the $SO(6)$ symmetry in the limit of $\l\to \l'$.
Under these assumption, the Nambu-Goldstone bosons are charged under the $SO(5)$ symmetry 
and are stable.  

For later convenience, let us solve the quantum modified constraint explicitly by,
\begin{eqnarray}
\label{eq:M0}
  M_{0} = \sqrt{ \Lambda_{\rm dyn}^{2}- \sum_{a=1-5} M_{a}^{2}}\ ,
\end{eqnarray}
and plug it into the effective superpotential in
Eq.\,(\ref{eq:SQQ2});
\begin{eqnarray}
\label{eq:Weff}
W_{\rm eff} &\simeq& \lambda \,\L_{\rm dyn}^{2} S_{0} 
- \sum_{a=1-5} \frac{\lambda}{2}\, S_{0}M_{a}^{2} 
+ \sum_{a=1-5}\lambda'\,\L_{\rm dyn}S_{a}M_{a} + O(M_{a}^{4})\ .
\end{eqnarray}
From this expression, we see that the SUSY breaking vacuum is given by,
\begin{eqnarray}
\label{eq:vac}
 F_{S_{0}} = \lambda \L_{\rm dyn}^{2}\
 ,\quad
  S_{a} = 0\ ,\quad
 M_{a} = 0\ .
\end{eqnarray}

\subsubsection{Mass spectrum of $S_0$ multiplet}
Around the vacuum in Eq.\,(\ref{eq:vac}), the tree-level potential in the $S_0$ direction 
is flat, and the masses of the $S_0$ multiplets can 
be significantly lighter than the dynamical scale.%
\footnote{The fermion components of $S_0$ corresponds to the 
spin one half component of the gravitino.}
As discussed in Ref.\,\cite{Ibe:2009dx}, this is a quite favorable
feature to account for the observed dark matter density.
The annihilation cross section of the Nambu-Goldstone dark matter 
with a mass in the TeV range is generically too small to explain the observed dark matter density
(see Ref.\,\cite{Ibe:2009dx} for general discussion).
In this model, however,  the annihilation cross section is enhanced
by a narrow resonance which is served by the flaton, a radial component of $S_0$.
It should be noted that, for that purpose, 
the R-symmetry must be broken so that the dark matters with 
R-charge zero annihilate through the flaton resonance with R-charge two.

In Ref.\,\cite{Ibe:2009dx}, we considered the model with spontaneous R-symmetry breaking.
The spontaneous R-symmetry breaking is caused by effects of 
higher dimensional operators of $S_{0}$ in the K\"ahler potential, 
\begin{eqnarray}
\label{eq:Kahler}
 K  &=& |S_{0}|^{2}+\frac{|S_{0}|^{4}}{4\Lambda_{4}^{2}}
 -\frac{|S_{0}|^{6}}{9\Lambda_{6}^{4}} + \cdots,
\end{eqnarray} 
where $\Lambda$'s denote the dimensionful parameters and the ellipsis
denotes the higher dimensional terms of $S_{0}$.  
The above K\"ahler potential provides an effective description of a quite
general class of the models with spontaneous breaking of the
R-symmetry breaking.  Especially, when the above K\"ahler potential
results from radiative corrections from physics at the scale
$\L_{\rm dyn}$, the dimensionful parameters are expected to be,
\begin{eqnarray}
\label{eq:dim4}
 \frac{1}{\L_{4}^{2}} = \frac{c_{4}^{2}}{16\pi^{2}}\frac{1}{\L_{\rm dyn}^{2}},
 \quad
  \frac{1}{\L_{6}^{4}} = \frac{c_{6}^{2}}{16\pi^{2}}\frac{1}{\L_{\rm dyn}^{4}},
\end{eqnarray}
where dimensionless coefficients $c_{4,6}$ are of the order of 
unity (see Ref.\,\cite{Ibe:2009dx} for an explicit model with R-symmetry breaking).
From this K\"ahler potential, the R-symmetry is spontaneously
broken by the vacuum expectation value of the scalar component of $S_{0}$;
\begin{eqnarray}
\label{eq:S0}
\vev{S_{0}} &=& \frac{1}{\sqrt{2}} \frac{\Lambda_{6}^{2}}{\Lambda_{4}}
= \frac{1}{\sqrt{2}}\frac{c_{4}}{c_{6}}\L_{\rm dyn}
= \frac{1}{\sqrt{2}}f_{R},
\end{eqnarray}
where we have introduced the R-symmetry breaking scale $f_{R} =
O(\L_{\rm dyn})$.

Now let us consider the masses of the scalar components of $S_0$.
At this vacuum, the scalar component of $S_{0}$ is decomposed into the
flaton $s$ and the R-axion $a$ by,
\begin{eqnarray}
  S_{0} = \frac{1}{\sqrt{2}} (f_{R}+s) e^{i a/f_{R}}.
\end{eqnarray}
Then, the mass of the flaton is given by,
\begin{eqnarray}
\label{eq:mflaton}
 m_{s} = 4\sqrt{2}\, 
\frac{\l\L_{\rm dyn}^{2}\L_{4}^{3}}{(4\L_{4}^{4}+ \L_{6}^{4})}
\simeq \sqrt{2} \frac{\l \L_{\rm dyn}^{2}}{\L_{4}}
\simeq \sqrt{2} \frac{c_{4}}{4\pi} \l \L_{\rm dyn},
\end{eqnarray}
where we have used $F_{S_{0}} = \l \L_{\rm dyn}^{2}$ and assumed
Eq.\,(\ref{eq:dim4}) with $c_{4}= c_{6}=O(1)$.  Therefore, the flaton
can be in the TeV range for $\l\sim 1$ and $c_{4}\sim 1$, which is a
crucial property for the flaton to make the narrow resonance
appropriate for the dark matter annihilation.%
\footnote{Even for large couplings, $\l$, $c_{4,6}$, there is a possibility 
that the model involves a ``light" flaton of a mass in the TeV range,
although those models are incalculable.}

The R-axion mass, on the other hand, can be much more suppressed,
since it is a pseudo-Goldstone boson of the spontaneous R-symmetry breaking. 
For example, when explicit breaking of the R-symmetry 
mainly comes from the constant term in the superpotential,%
\footnote{In this study, we assume that the messenger sector of the
  gauge mediation also respects the R-symmetry.  Otherwise, the
  radiative correction to the K\"ahler potential of $S_0$ from the
  messenger sector gives rise to the dominant contribution to the
  R-axion mass.  
  The R-breaking mass from the Higgs sector, on the other hand, 
  is smaller than the one in Eq.\,(\ref{eq:Rmass}), 
  even if the so-called $\mu$-term does not respect the R-symmetry. 
  } 
the R-axion acquires a small
mass\,\cite{Bagger:1994hh},
\begin{eqnarray}
\label{eq:Rmass}
m_{\mathrm{axion}}^2 \sim \frac{m_{3/2} F_{S_{0}} }{f_{R}}.
\end{eqnarray}
This mass is in a MeV range
for $f_R \simeq \sqrt{F_{S_0}}\simeq 30$\,TeV and $m_{3/2}\simeq 1$\,eV, for example.

The most interesting feature of the R-axion in the above mass range is that 
it mainly decays into an electron pair\,\cite{Goh:2008xz}.
Therefore, the dark matter which annihilates (or decays) into a pair of the R-axions 
ends up with the final states with four electrons.
In section\,\ref{sec:spectra}, we see that the four electron final states of 
the dark matter annihilation or decay are favorable to explain 
the positron/electron spectra.
Notice that there is a lower  limit on the decay constant $f_R$ of the 
axion-like particles which mainly decay into pairs of electrons 
from the beam-dump experiments\,\cite{Bergsma:1985qz};
\begin{eqnarray}
 f_R \gtrsim 10\,{\rm TeV} \times \left(\frac{m_a}{1\,{\rm MeV}}\right)^{1/2}\quad 
\mbox{for } 1\,{\rm MeV}\lesssim m_a \lesssim 100\,{\rm MeV}.
\end{eqnarray}
On the other hand, as we will see in section\,\ref{sec:density}, 
the right amount of dark matter requires $f_R \simeq 30$\,TeV.
Thus, the R-axion decaying into a electron pair in Nambu-Goldstone dark matter models
is consistent with the constraint from the beam-dump experiment
(see for example Ref.\,\cite{Kim:1986ax} for detailed constraints on the axion-like particles).

One may also consider explicit R-symmetry breaking by higher dimensional terms 
which are suppressed by some mass scale, $M_*$, such as a quartic term $S_0^4/M_*$.%
\footnote{Here, we assume $Z_{6R}$ symmetry instead of the continuous R-symmetry.
The main conclusion of this paper is not affected as long as the order of the 
discrete symmetry is high enough.}
With this breaking term, the R-axion mass in hundreds MeV range 
is realized for $f_R \simeq 30$\,TeV and $M_*\simeq 10^{16-17}$\,GeV.
In this case, the R-axion mainly decays into a pair of muons.
In the following analyses, we consider the R-axion of a mass in both 
a MeV range and  hundreds MeV range.

\subsubsection{Mass spectrum of $S_a$ and $M_a$ multiplets}
We now consider the spectrum of other light particles, $S_a$ and $M_a$.
From the superpotential Eq.\,(\ref{eq:Weff}), the fermion components of $S_a$ and $M_a$
obtain masses of $O(\l \L_{\rm dyn})$.
Scalar masses, on the other hand, comes from the scalar potential,
\begin{eqnarray}
\label{eq:scalarDM}
V &=& | \l' \L_{\rm dyn} \,M_{a}|^{2} + \left| \l \L_{\rm dyn}^{2} -
  \frac{\l}{2}M_{a}^{2}\right|^{2} 
+ | \l S_{0} M_{a} + \l'\L_{\rm dyn} S_{a} | ^{2},\cr
&=& \frac{1}{2}\left( (\l'^{2}-\l^{2})\L_{\rm dyn}^{2}
  +\frac{\l^{2}}{2} (s+f_{R})^{2}\right) x_{m}^{2}  
+ \frac{1}{2}\left( (\l'^{2}+\l^{2})\L_{\rm dyn}^{2} +\frac{\l^{2}}{2}
  (s+f_{R})^{2}\right) y_{m}^{2}\cr 
&&+\,\frac{1}{2}\l'^{2} \L_{\rm dyn}^{2} x_{s}^{2}+\frac{1}{2}\l'^{2}
\L_{\rm dyn}^{2} y_{s}^{2}  
+ \frac{\l \l'}{\sqrt{2}} \L_{\rm dyn} (s+f_{R})\, x_{m} x_{s} +
\frac{\l \l'}{\sqrt{2}} \L_{\rm dyn} (s+f_{R})\, y_{m} y_{s}\cr 
&&+ \,\l^2\L_{\rm dyn}^4 + \frac{\l^2}{2^4}(x_m^2+y_m^2)^2,
\end{eqnarray}
where we have decomposed the scalars by,
\begin{eqnarray}
  S_{a} &=& \frac{1}{\sqrt 2} (x_{s}+i\, y_{s})e^{i a/f_{R}}, \cr
  M_{a} &=& \frac{1}{\sqrt 2} (x_{m}+i\, y_{m}),
\end{eqnarray}
and suppressed the index of $SO(5)$.
Notice that the R-axion does not show up in the scalar interactions in
this basis, and it only appears in the derivative couplings. 

From the above potential, we find that the pseudo-Nambu--Goldstone mode resides
not in $(y_{m}, y_{s})$ but in $(x_{m}, x_{s})$ and
the eigenvalues of the squared mass matrix of  $(x_{m}, x_{s})$ are given by,
\begin{eqnarray}
 m_{\phi}^{2} &=&\frac{1}{2} \left(\tr M^{2}- \sqrt{ (\tr
     M^{2})^{2}-4\det M^{2}}  \right)=\frac{\det M^{2}}{m_{H}^{2}}
     \simeq  \frac{ (\l'^2-\l^2) \L_{\rm dyn}^4}{  \L_{\rm dyn}^2 + 2f_R^2},\\ 
 m_{H}^{2} &=& \frac{1}{2} \left(\tr M^{2}+ \sqrt{ (\tr
     M^{2})^{2}-4\det M^{2}}  \right)
      \simeq \l^2 \left(\L_{\rm dyn}^2 + \vev{S_0}^2 \right),\\ 
 \tr M^{2} &=& (2\l'^{2}-\l^{2}) \L_{\rm dyn}^{2} + \l^{2}\vev{S_{0}}^{2},\\
  \det M^{2} &=& \l'^{2}( \l'^{2}-\l^{2})\L_{\rm dyn}^{4},
\end{eqnarray}
where the rightmost expressions of $m_{\phi,H}^2$ are valid for $\l \simeq \l'$.
The eigen mode $\phi$ corresponds to the pseudo-Nambu-Gladstone boson
of the approximate $SO(6)$ symmetry, which becomes massless in the limit of 
an exact $SO(6)$, i.e. $\l\to\l'$.
The masses of $(y_m,y_s)$ are, on the other hand, of $O(\l \L_{\rm dyn})$.

The R-axion interactions only appear in the kinetic terms.  In the current
basis, the R-axion interactions come from the kinetic
terms of $S_{0}$ and $S_{a}$,
\begin{eqnarray}
\label{eq:kin}
{ \cal L } = \frac{1}{2} (\partial a)^{2}\left( 1 + \frac{s}{f_{R}} \right)^{2}
+ \frac{1}{2 f_{R}^{2} } (\partial a)^{2}(x_{s}^{2}+y_{s}^{2} )
+ \frac{1}{f_{R}} \partial_{\m} a (x_{s}\partial^{\m} y_{s}  - y_{s}\partial^{\m} x_{s}).
\end{eqnarray}

Altogether, in the Nambu--Goldstone dark matter scenario ({\it i.e.}\/, $\l'-\l
\ll 1$), the light particle sector below $O(10)$\,TeV consists of the dark matter and the
flaton in the TeV range, and the gravitino and the R-axion with much
smaller masses, while the other components of $S_{a}$ and $M_{a}$
have masses of the order of the SUSY breaking scale, $\l^{1/2} \L_{\rm dyn}$.  The most
relevant terms for the dark matter annihilation are, then, given by,
\begin{eqnarray}
\label{eq:int}
{\cal L}_{\rm int} = \frac{\l^{2}}{2} f_{R} \frac{m_{\phi}^{2}} {m_{H}^{2}-m_{\phi}^{2}}\, s\, \phi^{2}
+  \frac{1}{2} (\partial a)^{2}\left( 1 + \frac{s}{f_{R}} \right)^{2},
\end{eqnarray}
where the first term comes from the scalar potential in
Eq.\,(\ref{eq:scalarDM}), while the second term comes from
Eq.\,(\ref{eq:kin}).

\subsubsection{Flaton decay}\label{sec:flaton}
Now let us discuss the decay of flaton which is important
to estimate the dark matter annihilation via the s-channel
exchange of the flaton.
First, we consider the decay mode into a pair of the R-axions.  
The
relevant interactions for the decay come form the first term in
Eq.\,(\ref{eq:int}), and the decay rate into a pair of the R-axion is
given by,
\begin{eqnarray}
\label{eq:saa}
\G_{s\to aa} /m_s= \frac{1}{32\pi} \frac{m_{s}^{2}}{f_{R}^{2}} \simeq 4 \times 10^{-4}
\left(\frac{m_s}{6\,\rm TeV}\right)^2
\left(\frac{30\,\rm TeV}{f_R}\right)^2,
\end{eqnarray}
where we have neglected the mass of the R-axion and taken the final
state velocity to be $\b_{f}=1$.  

Next, we consider the flaton decay into a pair of the dark matter.
The relevant interaction term is given in Eq.\,(\ref{eq:int}) and the
resultant decay rate is given by,
\begin{eqnarray}
 \G_{s\to \phi\phi}/m_s = \frac{\beta_{\phi}}{32\pi}\frac{\l^{4}f_{R}^{2}}{m_{s}^{2}} \left(\frac{m_{\phi}^{2}}{m_{H}^{2}-m_{\phi}^{2}} \right)^{2},
\end{eqnarray}
where $\beta_{\phi}$ denotes the size of the velocity of the dark
matter.  Notice that the value of $\G_{s\to \phi\phi}/\beta_{\phi}$ is
well-defined even in the unphysical region, {\it i.e.}\/, $2 m_{\phi}> m_{s}$.
As a result, we find that  $\G_{s\to \phi\phi}/\beta_{\phi}$ is suppressed compared
with $\G_{s\to aa}$,
\begin{eqnarray}
 \G_{s\to \phi\phi}/m_s \simeq \frac{\beta_{\phi}}{512\pi} \left(\frac{\l f_{R}}{m_{H}}\right)^{4} 
 \frac{m_{s}^{2}}{f_{R}^{2}}
   \simeq \frac{\beta_{\phi}}{512\pi} \left(\frac{f_{R}^2}{\L_{\rm dyn}^2+2f_R^2}\right)^{2} 
 \frac{m_{s}^{2}}{f_{R}^{2}},
\end{eqnarray}
where we have used $m_{\phi}\simeq m_{s}/2$ and $m_{H}\gg m_{\phi}$.

As we will see in section\,\ref{sec:ssm}, the flaton also decays into a pair of the SSM
particles, which are roughly suppressed by the mass ratio squared 
of the SSM fields and the flaton compared with the R-axion mode.
Thus, to determine the dark matter density, it is good enough to consider 
the decay into the R-axion.
Since we are mainly interested in the R-axion of a mass in a MeV or hundreds MeV range,
the R-axion in the final state eventually decays into pairs of electrons or muons.
Therefore, the dark matter annihilation ends up with the final state with four electrons or four muons.

Put it all together, we obtain the flaton decay width at $E_{\rm
  CM}\simeq m_{s}$,
\begin{eqnarray}
 \G_{s} (E_{\rm CM}) = \G_{s\to aa} + \G_{s\to \phi\phi} + \cdots.
\end{eqnarray}
where $E_{\rm CM}>2 m_{\phi}$, and the ellipses denotes the modes into
the SSM particles (see section\,\ref{sec:ssm}).  In the
following analysis, we approximate the above decay rate by,
\begin{eqnarray}
 \G_{s} (E_{\rm CM}) \simeq \G_{s}(m_{s})\simeq \G_{s\to aa},
\end{eqnarray}
since the other modes are subdominant for $E_{\rm CM}\simeq m_{s}$.

\subsection{Symmetry and Stability of Dark Matter}
So far, we have assumed that the $SO(5)$ global symmetry out
of the maximal $SO(6)$ is exact and the dark matter is completely stable.
This symmetry has been imposed by hand to make the dark matter stable.
Although there is nothing wrong with this assumption,
it is more attractive if the stability of the dark matter is assured by
symmetries which are imposed by other reasons than the stability 
of the dark matter.
In this section, we show that the present model can have 
such an accidental symmetry by which the stability of the dark matter is achieved.
Once the stability is assured by an accidental symmetry,
the dark matter decays only through higher dimensional interactions which
do not respect the accidental symmetry.
As we will see in the following sections, the lifetime of the dark matter is long enough,
and furthermore, the lifetime can be in an appropriate range to explain the 
observed positron/electron excesses.

\subsubsection{Stability and accidental symmetry}
In the SUSY breaking model discussed above, 
we may  take a $U(1)$ subgroup of 
the global $SU(4)$ symmetry 
a gauge symmetry.
As discussed in Ref.\,\cite{Ibe:2009dx},
the SUSY breaking model 
with such a $U(1)$ gauge symmetry ensures spontaneous R-symmetry breaking 
in certain parameter space (see also Ref.\,\cite{Dine:2006xt}).%
\footnote{The introduction of the $U(1)$ gauge symmetry
solves the domain wall problem in the original vector-like SUSY breaking model
in Ref.\,\cite{Izawa:1996pk}. }

We assign the $U(1)$ charges to the fundamental fields; 
 $Q_{1}(1/2)$,  $Q_{2}(1/2)$,  $Q_{3}(-1/2)$ and $Q_{4}(-1/2)$,
 which corresponds to;
 $M_{12}(1)$,  $M_{34}(-1)$, while other mesons are neutral.
The charges of the singlets $S$'s are assigned so that the interaction in Eq.\,(\ref{eq:SQQ})
is consistent with the $U(1)$ symmetry.
In this case, the low-energy effective superpotential of the gauged IYIT model is
given by%
\begin{eqnarray}
W = \l S_+ M_- + \l S_- M_+ + \l_a S_a M_a\ , 
\end{eqnarray}
with the quantum constraint $2M_+M_- +M_aM_a =\L_{\rm dyn}^2 $.
Here, the subscript 
$a$ runs $a=1-4$, and $M_{\pm}$ corresponds to $M_{12}$ and $M_{34}$, respectively.
For $\l < \l_a$, the quantum constraint is satisfied by $2 M_+M_- = \L_{\rm dyn}^2$,
and the $U(1)$ gauge symmetry is broken down spontaneously.

Notably, with just the introduction of the $U(1)$ gauge symmetry, 
the model now has an accidental symmetry 
which assures the stability of the dark matter
under the gauge symmetries and the R-symmetry.
Here, let us remind ourselves that none of these interactions
are imposed to assure the stability of the dark matter.

Under the above symmetries,
the lowest dimensional interactions in the superpotential are those in Eq.\,(\ref{eq:SQQ}),
and the second lowest dimensional interactions are suppressed by some
high mass scale $M_{*}^2$,
such as terms in the superpotential, $SQ^4/M_*^2$.
Here, we are assuming the R-charge assignment;  $S(2)$ and $Q(0)$ (see discussion at the end of this section).
As a result, 
the model possesses an accidental global $U(1)$ symmetry, at the energy scale much lower than $M_*$, with the charge assignment; $S(2)$ and $Q(-1)$, which is in terms of the low energy fields; 
$S(2)$ and $M(-2)$.
This symmetry is broken down to a discrete $Z_4$ symmetry 
by an anomaly to the $SP(1)$ dynamics.
Altogether, in the low energy theory, the model has an accidental $Z_4$
symmetry (which is effectively a $Z_2$ symmetry in terms of the low energy 
effective fields, $S$'s and $M$'s).

After spontaneous SUSY breaking, 
the R-symmetry and the $U(1)$ gauge symmetry are broken spontaneously
by the vacuum condition $2M_+M_-=\L_{\rm dyn}^2$ and $S_+S_-\neq 0$.
Then, an appropriate combination of the accidental $Z_4$ symmetry and 
the $U(1)$ gauge symmetry remains unbroken at the above vacuum.
Concretely, a rotation under the $U(1)$ gauge symmetry 
by an angle $\pi$ on $M_{\pm}$ is cancelled 
by a change of signs of $M_{\pm}$ under the $Z_4$ symmetry,
while the dark matters change signs under this unbroken combination.
Therefore, there is an accidental $Z_4$ symmetry at the vacuum,
under which the dark matter changes sign.

Put it all together,  in the SUSY breaking model with the $U(1)$ gauge symmetry,
the dark matter stability is assured by the accidental $Z_4$ symmetry
which results from the gauge symmetries and the R-symmetry.
The lowest dimensional interactions which  break
the $Z_4$ symmetry are suppressed by some high mass scale $M_*^2$,
and hence, the lifetime of the dark matter decaying via the symmetry breaking 
interactions is much longer than the age of the universe.
Furthermore, as we will see below, the lifetime of the dark matter by the decay 
via the lowest dimensional symmetry breaking terms is in an appropriate range 
to explain the observed positron/electron spectra.

\subsubsection{Decaying dark matter}
The lowest dimensional interactions which violate the accidental $Z_4$ symmetry
are, for example, given by the following terms in K\"ahler potential,
\begin{eqnarray}
\label{eq:breaking}
 K = c_0 \sum Q_i Q_j = c\, \L_{\rm dyn} \sum_{a=1-4}  M_a,
\end{eqnarray}
which  respect gauge symmetries as well as the R-symmetry.
Here, we have omitted flavor indices of dimensionless coefficients, $c_0$ and $c$,
and rewritten the operators in terms of the low energy fields in the rightmost expression.
Notice that the above operators have physical effects only through supergravity,
and hence, the effects are suppressed by the Planck scale 
$M_*\simeq M_{\rm PL}\simeq 2.4 \times 10^{18}$\,GeV.

In supergravity, the above terms result in mass mixings between the flaton and the Nambu-Goldstone dark matters,
\begin{eqnarray}
 \delta V= \frac{ c  }{\sqrt{2}}
 \left(\frac{\L_{\rm dyn}}{M_{\rm PL}} \right)^2
 \left(\frac{f_R}{\L_{\rm dyn}}\right)\left(\l \L_{\rm dyn}\right)^2\, x_m s
 = \delta m^2\, x_m s ,
\end{eqnarray}
and the mixing angles are given by,
\begin{eqnarray}
\label{eq:mix}
 \varepsilon \simeq \frac{ \delta m^2}{m_s^2 - m_\phi^2} = \frac{\delta m^2}{3 m_\phi^2},
  \end{eqnarray}
where we have omitted flavor indices of the dark matters again and used $m_s \simeq 2 m_\phi$
in the final expression.
With these mixings, the dark matters decay in a similar way of the flaton discussed in section\,\ref{sec:flaton},%
\footnote{With the above symmetry breaking term, the dark matter decays into the 
MSSM particles with the same suppression factor $M_{PL}^2$.
The branching ratios of those modes, however, are of $O(m_{\phi}^6/\L_{\rm dyn}^6)$
and highly suppressed.
}
and the main decay mode is that into the R-axion pair and the lifetime is given by,
\begin{eqnarray}
\tau_{\rm DM} \simeq 3 \times 10^{27}\,{\rm sec}\times
\left(\frac{1}{c}\right)^2
\left(\frac{3}{\l}\right)^4
\left(\frac{30\,\rm TeV}{\L_{\rm dyn}}\right)^6
\left(\frac{m_{\phi}}{6\,\rm TeV}\right).
\end{eqnarray}
Here, we have taken $f_R \simeq \L_{\rm dyn}$ for simplicity. 
As a result, the lifetime of the dark matter is much longer than the age of the universe.

Before closing this section, let us comment on how accidental the above ``accidental" symmetry is.
In the above discussion, we have assigned R-charges $S(2)$ and $Q(0)$,
so that the  R-symmetry forbids the $Z_4$ breaking terms 
suppressed by a factor of $M_*^{-1}$,  while
allowing the breaking terms suppressed by $M_*^{-2}$ as given in Eq.\,(\ref{eq:breaking}).
With this charge assignment, however, linear terms of $S_a$ in a superpotential 
are still consistent with all the other symmetries than the ``accidental" $Z_4$, while they break the ``accidental" $Z_4$ symmetry.
Thus, strictly speaking, the $Z_4$ symmetry is hard broken, and hence,
it cannot be an accidental symmetry.
To avoid this problem, one may  consider another charge assignment of the R-symmetry;
$S(2/5)$ and $Q(4/5)$.
Under this charge assignment, the linear terms of $S$'s are forbidden and the
lowest $Z_4$ breaking terms are given by $S^5/M_*^2$, which result in 
similar mass mixings between the dark matter and the flaton given in Eq.\,(\ref{eq:mix}).
In this way, we may have a truly accidental $Z_4$ symmetry which is the outcome
of the gauge symmetries and the R-symmetry.%
\footnote{
As an alternative solution, we may forbid the linear terms of $S$'s 
by assuming a conformal symmetry in the limit of the vanishing gravitational interactions.
Once the linear terms of $S$'s are forbidden at the tree-level, they are not generated 
by any radiative corrections.}

\subsection{Dark matter density}\label{sec:density}
In this section, we discuss the parameter space which reproduces 
the observed dark matter density by the Nambu-Goldstone dark matter.
\subsubsection{Annihilation cross section via flaton resonance}
As derived in Ref.\,\cite{Ibe:2009dx}, the annihilation cross section of the dark matter into a pair of the R-axion via the flaton resonance  is given by,
\begin{eqnarray}
\label{eq:DMCS}
\s v_{\rm rel} = 
\frac{v_{\rm rel}}{32\pi} \frac{\b_{f}}{\b_{\phi}}
\left( \frac{m_{\phi}^{2}}{m_{H}^{2}-m_{\phi}^{2}}\right)^{2}
\frac{\l^{4}E_{\rm CM}^{2}} { (E_{\rm CM}^{2}-m_{s}^{2})^{2} + m_{s}^{2}\G_{s}^{2}}
\simeq 
\frac{1}{64\pi} 
\left( \frac{m_{\phi}^2}{\L_{\rm dyn}^2+2f_R^2}\right)^{2} \frac{1}{m_{\phi}^{2}}
\frac{1}{ (\delta + v_{\rm rel}^{2}/4)^{2} + \g_{s}^{2}},
\end{eqnarray}
where the $\G_{s}$ and $\b_{\phi}$ are defined at $E_{\rm CM}>2m_{\phi}$.
In the final expression, we have used the non-relativistic approximation,
\begin{eqnarray}
  E_{\rm CM}^{2} = 4 m_{\phi}^{2} + m_{\phi}^{2} v_{\rm rel}^{2} \ ,
  \quad
  \beta_{\phi} = \sqrt{1- 4 m_\phi^2/E_{\rm CM}^2} \ ,
\end{eqnarray}
and introduced parameters $\d$ and $\g_{s}$ by,
\begin{eqnarray}
m_{s}^{2}=4m_{\phi}^{2}(1-\delta),\quad \g_{s} = \G_{s}/m_{s}.
\end{eqnarray}
Notice that the cross section does not depend on the parameter $\lambda$ explicitly.

\subsubsection{Required cross section}
In the presence of the narrow resonance, 
the thermal history of the dark  matter density is drastically changed from the usual thermal relic
density without the resonance\,\cite{Griest:1990kh,Gondolo:1990dk}. 
As a result, the required annihilation cross section to account for 
the observed dark matter density\,\cite{Komatsu:2008hk}
is different from the one for the non-resonant annihilation cross section,
\begin{eqnarray}
 \vev{\s v_{\rm rel}} \sim 10^{-9}\,{\rm GeV}.
\end{eqnarray}
Instead, in terms of the annihilation cross
section at the zero temperature, the required annihilation cross
section to obtain the correct abundance is given by\,\cite{Ibe:2008ye},
\begin{eqnarray}
\label{eq:csreq}
 \vev{\s v_{\rm rel}}|_{T = 0} \sim 10^{-9}\,{\rm GeV}^{-2} \times \frac{x_{b}}{x_{f}}.
\end{eqnarray}
Here $x_{f}\simeq 30$ denotes the freeze-out parameter of the usual
(non-resonant) thermal freeze-out history, while the effective freeze-out parameter
$x_{b}$ is defined by,
\begin{eqnarray}
\frac{1}{x_{b}} \simeq \frac{1}{\vev{\s v_{\rm rel}}|_{T=0} }
\int_{x_{f}}^{\infty}\frac{\vev{\s v_{\rm rel}}}{x^{2}} \,dx.
\end{eqnarray}

For a well tuned and very narrow flaton, i.e. $|\d|,\g_s\ll1$, 
we found that the effective freeze-out parameter is fairly approximated by,
\begin{eqnarray}
 x_b^{-1} \simeq\frac{\d^2 + \g_s^2}{\g_s} 
 \left( 
 \frac{\pi}{2} - \arctan\left[\frac{\delta}{\gamma_s}\right] 
 \right).
  \end{eqnarray}
From this expression, we find  
the asymptotic behaviors of the effective freeze-out parameter as follows.
\begin{itemize}
\item Unphysical pole ($\d>0$), $\d \ll \gamma_s\ll 10^{-1}$
\begin{eqnarray}
\label{eq:xb1}
 x_b^{-1} \simeq  \pi\, \gamma_s\ .
\end{eqnarray}
\item Unphysical pole ($\d>0$), $\g_s \ll \d\ll 10^{-1}$
\begin{eqnarray}
\label{eq:xb2}
 x_b^{-1} \simeq 2\,\delta \ .
\end{eqnarray}
\item Physical pole ($\d<0$), $|\d| \ll \gamma_s\ll 10^{-1}$
\begin{eqnarray}
\label{eq:xb3}
 x_b^{-1} \simeq \pi\,\gamma_s \ .  
\end{eqnarray}
\item Physical pole ($\d<0$), $\g_s \ll |\d|\ll 10^{-1}$
\begin{eqnarray}
\label{eq:xb4}
 x_b^{-1} \simeq 2\pi\, \frac{\d^2}{\g_s}\ .
\end{eqnarray}
\end{itemize}
Here, we have used $\arctan(1/x) = \pi/2 - \arctan(x)$ for $x>0$ in Eq.\,(\ref{eq:xb2}),
and $\arctan(1/x) = -\pi/2 - \arctan(x)$ for $x<0$ in Eq.\,(\ref{eq:xb2}).
Notice that the Breit--Wigner enhancement is realized for the first three cases, that is,
\begin{eqnarray}
  \vev{\s v_{\rm rel}}|_{T = 0} \gg 10^{-9}\,{\rm GeV}^{-2}\ .
\end{eqnarray}
while the late time cross section can be smaller than the non-resonant annihilation
in the final case, i.e.,
\begin{eqnarray}
  \vev{\s v_{\rm rel}}|_{T = 0} \ll 10^{-9}\,{\rm GeV}^{-2} \ .
\end{eqnarray}

\subsubsection{Constraints on SUSY breaking scale}
Now let us compare the cross section of the Nambu-Goldstone dark matter
via the narrow flaton resonance in Eq.\,(\ref{eq:DMCS}) and the required cross section
in Eq.\,(\ref{eq:csreq}).
In the limit of 
$|\d|\ll \g_s$ (see Eqs.\,(\ref{eq:xb1}) and (\ref{eq:xb3})),
the required annihilation cross section in Eq.\,({\ref{eq:csreq}}) is reduced to,
\begin{eqnarray}
\label{eq:wmap3}
\vev{ \sigma v_{\rm rel}}|_{T=0} \sim 10^{-9}\,{\rm GeV}^{-2} \times 
\frac{1}{x_{f}\pi\gamma_{s}} \simeq 8\times 10^{-9}\,{\rm GeV}^{-2} \times 
\frac{f_R^2}{x_f\,m_\phi^2} \ .
\end{eqnarray}
Here, we have used the flaton decay rate in Eq.\,(\ref{eq:saa}).
In the same limit, the predicted cross section in Eq.\,(\ref{eq:DMCS}) is 
\begin{eqnarray}
\label{eq:predicted}
\vev{ \sigma v_{\rm rel}}|_{T=0} \simeq \frac{1} {64\pi}
\left( \frac{m_{\phi}^2}{\L_{\rm dyn}^2+2f_R^2}\right)^{2} \frac{1}{m_{\phi}^{2}}
\frac{1}{  \g_{s}^{2}}
\simeq \frac{\pi f_R^4}{(\L_{\rm dyn}^2 + 2f_R^2)^2m_\phi^2},
\end{eqnarray}
where, again, we have used $\gamma_s$ in Eq.\,(\ref{eq:saa}).

First, we consider the case of the unphysical pole, $\d>0$.
In this case, the require cross section decreases in $\delta^{-1}$ for a larger $\d$
and in a region of $\d>\g_s$, 
while the predicted cross section decreases in $\delta^{-2}$. 
Therefore, in order for the predicted cross section meets the required value, 
the predicted cross section in the limit of $\d\ll \g_s$ must be larger than the required one, {\it i.e.},
\begin{eqnarray}
 \frac{\pi f_R^4}{(\L_{\rm dyn}^2 + 2f_R^2)^2m_\phi^2} >  8\times 10^{-9}\,{\rm GeV}^{-2} \times 
\frac{f_R^2}{x_f\,m_\phi^2}.
\end{eqnarray}
Thus, for $x_f \simeq 30$, we find a constraint on $f_R$ and $\L_{\rm dyn}$,
\begin{eqnarray}
\label{eq:constraint}
 f_R  \lesssim\, 50\,{\rm TeV} \times \left(1 +\frac{\L_{\rm dyn}^2}{2f_R^2} \right)^{-1}.
\end{eqnarray}
Notice that this bound is independent of the mass of the dark matter.

\begin{center}
\begin{figure}[t]
 \begin{minipage}{.45\linewidth}
  \includegraphics[width=\linewidth]{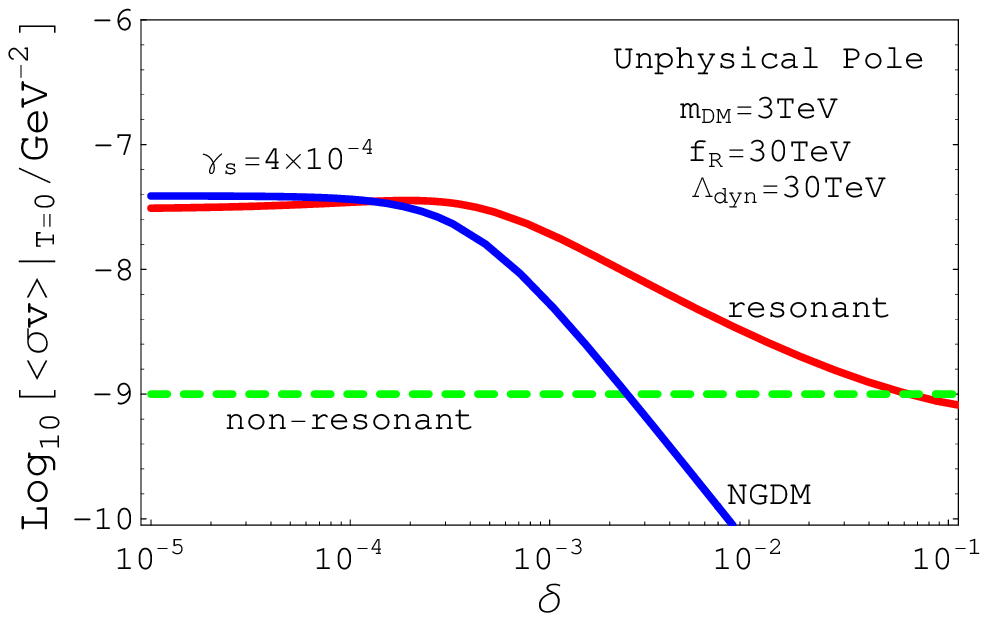}
 \end{minipage}
 \begin{minipage}{.45\linewidth}
  \includegraphics[width=1.0\linewidth]{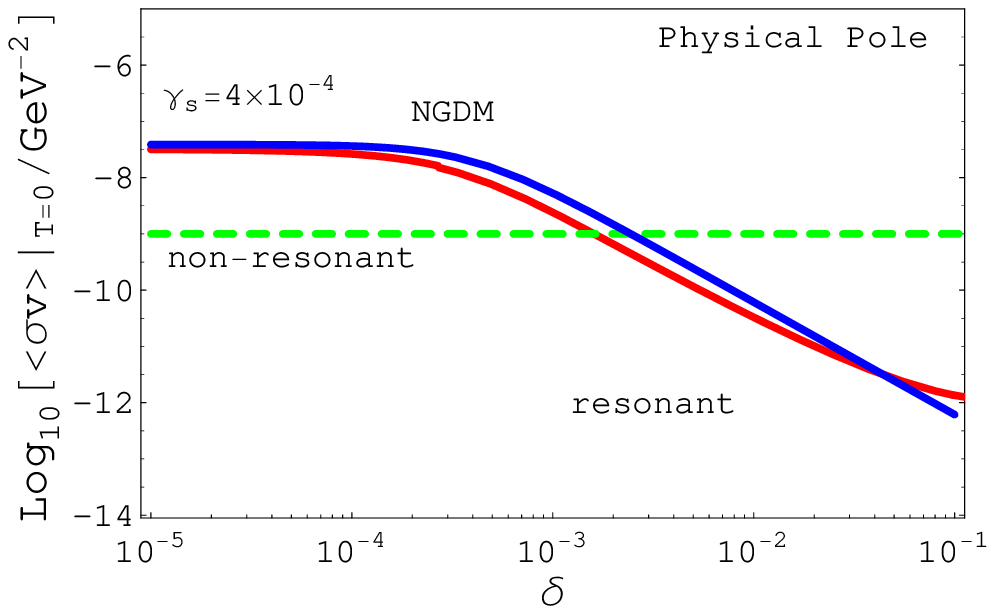}
 \end{minipage}
 \caption{Left) The $\delta$ dependence of the required annihilation
   cross section at zero temperature from the observed dark matter
   density in the case of the unphysical pole (red line)
  for $\g_{s}=4\times10^{-4}$. 
   $m_{\phi}=3$\,TeV, $\L_{\rm dyn}=30$\,TeV and $f_R = 30$\,TeV,
   which corresponds to 
   The green line shows the   required annihilation cross section in the usual thermal history.
   The blue line shows the predicted annihilation cross section for
   $m_{\phi}=3$\,TeV, $\L_{\rm dyn}=30$\,TeV and $f_R = 30$\,TeV,
   which corresponds to $\g_s = 4\times 10^{-4}$.
      Right) The required
   annihilation cross section in the case of the physical pole.  
   In the figures, we have assumed $x_f=30$.}
\label{fig:CS}
\end{figure}
\end{center}

In Fig.\,\ref{fig:CS}, we showed the comparison between the predicted 
and the required cross sections  for the unphysical pole for a given parameter set (Left).
Here, we numerically solved the Boltzmann equation of the dark matter density. 
The figure shows that the predicted cross section satisfy the required value
at $\delta \simeq 10^{-4}$ for $\L_{\rm dyn} = f_R = 30$\,TeV.
For  larger values of $\L_{\rm dyn}$  and $f_R$, the predicted cross section
is always smaller than the required one.

With the above constraint on $\L_{\rm dyn}$ and $f_R$, we may 
derive an upper bound on the effective boost factor in the Breit--Wigner enhancement
in the present model.
The effective boost factor is given by, 
\begin{eqnarray}
 {\rm BF} = \frac{x_b}{x_f},
\end{eqnarray}
 and, for a give value of $\gamma_s$, the factor is constraint from above by,
 \begin{eqnarray}
{\rm BF} \lesssim  \frac{1}{x_f\,\pi\,\g_s}=\frac{8 f_R^2}{x_f\,m_\phi^2}
\simeq 30 \times
\left(\frac{30}{x_f}\right)
\left(\frac{f_R}{30\,{\rm TeV}}\right)^2
\left(\frac{3\,{\rm TeV}}{m_\phi}\right)^2.
\end{eqnarray}
As a result, we found that the effective boost factor cannot be significantly
larger than $O(10)$ in this Nambu-Goldstone dark matter model.%
\footnote{See Ref\,.\cite{CyrRacine:2009yn} for protohalo constraints on the effective boost factor of the dark matter annihilation.}

Second, we consider the case with the physical pole, $\d<0$.
In this case, both the predicted and the required cross section decreases in $\d^{-2}$
in a region of $|\d|>\g_s$.
This means that the predicted cross section  satisfies the required value only for a special value
of $\L_{\rm dyn}$ and $f_R$, i.e.
\begin{eqnarray}
\label{eq:constraint2}
 f_R  \sim\, 50\,{\rm TeV} \times \left(1 +\frac{\L_{\rm dyn}^2}{2f_R^2} \right)^{-1}.
\end{eqnarray}
Inversely, once $\L_{\rm dyn}$ and $f_R$ take these values,
the predicted dark matter density is consistent with observation for 
a wide range of $\delta$.
In the right panel of Fig.\,\ref{fig:CS}, we compared the predicted and the required cross sections.
The figure shows that the predicted cross section is consistent with the required cross section
in a wide range of $\delta$ for $\L_{\rm dyn}\simeq f_R \simeq 30$\,TeV.


Put it altogether, the both scenarios require the dynamical scale and the 
decay constant around $\L_{\rm dyn}\simeq f_R \simeq 30$\,TeV.
From the analysis in the previous section\,\ref{sec:model}, 
these leads to the SUSY breaking scale $F\simeq \l^{1/2} \times 30$\,TeV.
It should be noted, however, that there are $O(1)$ ambiguities 
associated with the strong dynamics
to relate the dynamical scale and the SUSY breaking scale, 
and hence, 
the SUSY breaking scale can be slightly different from this naive expectation.

\section{Cosmic Ray Spectra in Nambu-Goldstone Dark Matter}\label{sec:spectra}
In this section, we discuss the cosmic ray spectra in the Nambu-Goldstone 
dark matter scenario.
As we have seen, the right amount of the dark matter density is achieved by 
the annihilation process via the flaton resonance in the early universe.
The cosmic ray spectra, on the other hand, depend on how the dark matters behave 
at the later time of the universe.
In the Nambu-Goldstone dark matter scenario, we have two options for the behaviors
at the later time. 
One option is the scenario with the enhanced annihilation and the other is 
the decaying dark matter.
In the case of the annihilating dark matter, the annihilation cross section at the later 
universe is important.
As we have seen, the unphysical flaton pole results in the Breit-Wigner enhancement,
which greatly reduces the necessity of the astrophysical boost of the annihilation 
cross section.
In the case of the decaying dark matter, the cosmic ray spectra can be explained 
regardless of the annihilation cross section, and hence,
scenario works well for both the physical and unphysical pole scenarios.

In the followings, we consider both options.
As we will see, both options can fit the observed positron/electron spectrum.
We further discuss the distinctive prediction on the gamma-ray spectrum, which
comes from the finite annihilation/decay rates into the SSM particles.

\subsubsection{The electron/positron excesses from the Nambu-Goldstone dark matter}
We first consider the electron/positron spectra from the decay/annihilation
of the Nambu-Goldstone dark matter with the 
final states consist of a pair of the R-axions which subsequently decay into pairs of
electrons or muons.%
\footnote{Once the muon modes of the R-axion is open, the electron mode
is negligible, and hence, we do not need to consider the final state with 
the mixed leptons $e^+e^-\m^+\m^-$.}
The detailed parameter sets used for the decay/annihilation of the dark matter
are given in Table\,\ref{tb:DM}.

\begin{table}[tbp]
\caption{Setup of the DM decay and annihilation}
\label{tb:DM}
\begin{tabular}{|c|c|c|c|c|}
\hline
Mode& $m_{\rm DM}$ &$m_a$ & Decay of $a$& Lifetime/Boost factor   \\ \hline  \hline
Decay& 5 TeV & 250 MeV &$\mu^+\mu^-$ & $1.5 \times 10^{26}$ sec       \\ \hline
Decay&1.5 TeV & 5 MeV & $e^+e^-$   & $5 \times 10^{26}$ sec   \\ \hline
Annihilation& 2.5 TeV & 250 MeV& $\mu^+\mu^-$  & 1500   \\ \hline
Annihilation& 750 GeV & 5 MeV  &$e^+ e-$& 150   \\ \hline
\end{tabular}
\end{table}

\begin{figure}[t!]
\begin{tabular}{lr}
\begin{minipage}{0.5\hsize}
\begin{center}
\epsfig{file=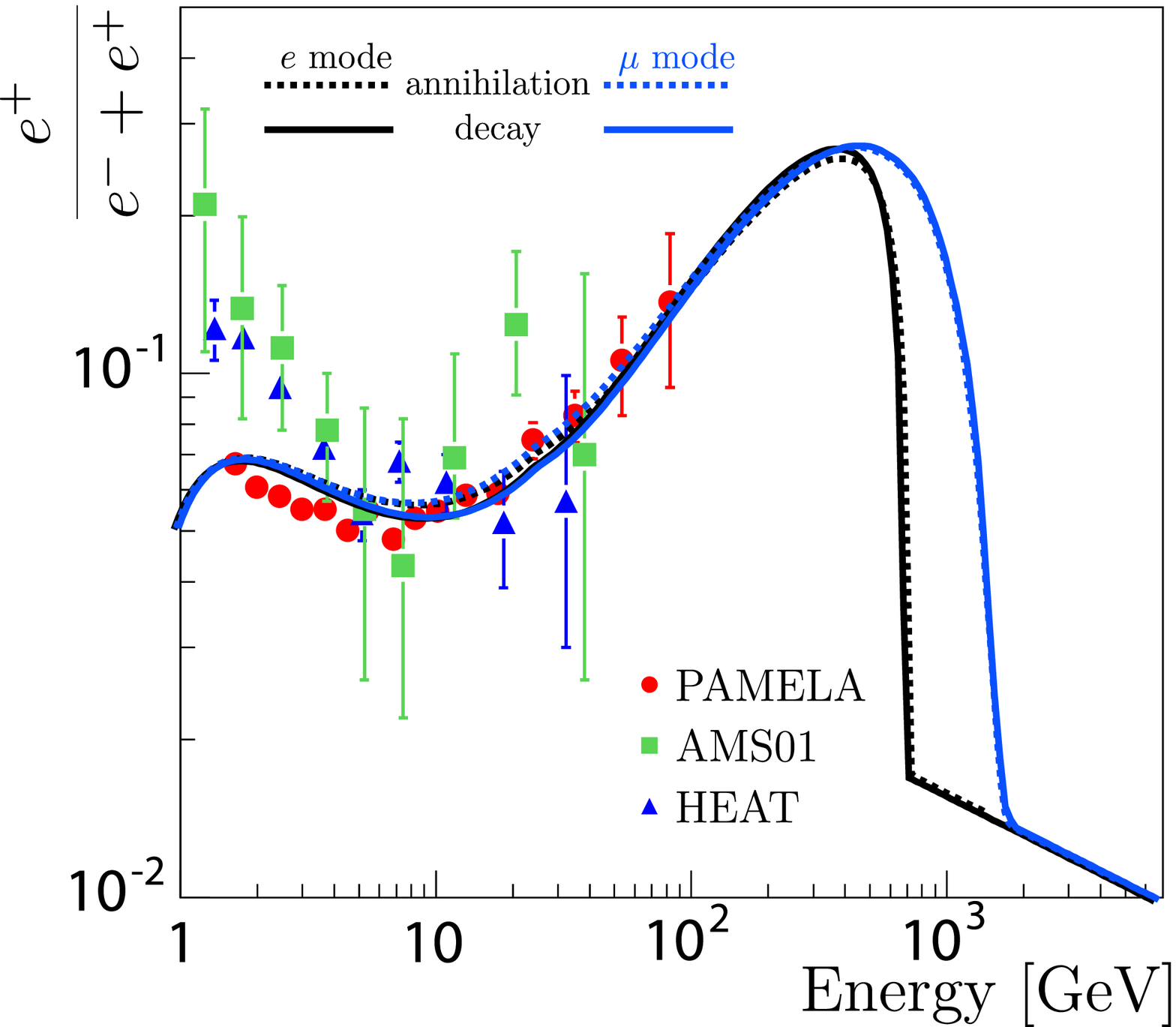 ,scale=.42,clip}\\
(a)
\end{center}
\end{minipage}
\begin{minipage}{0.5\hsize}
\begin{center}
\epsfig{file=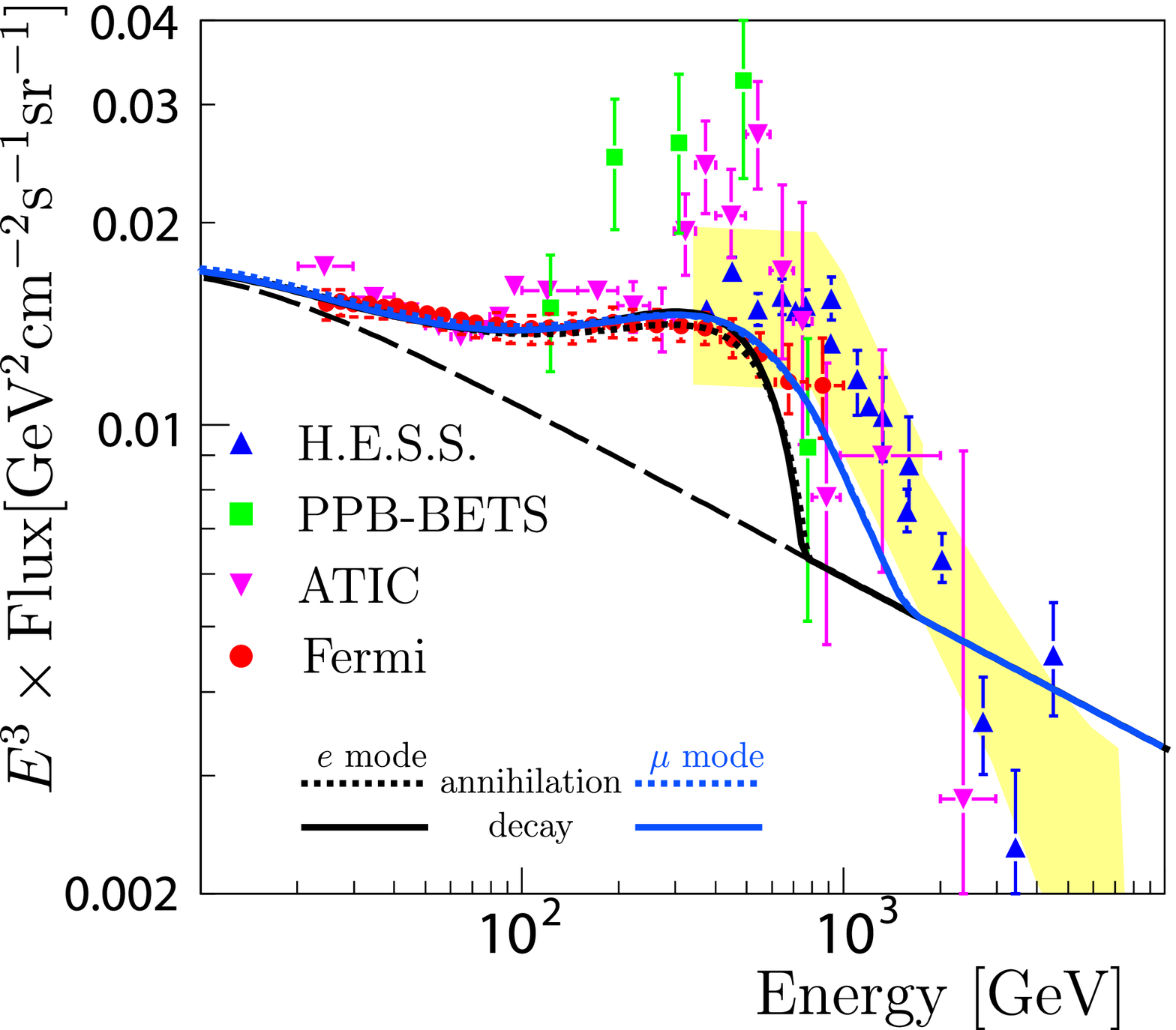 ,scale=.42,clip}\\
(b)
\end{center}
\end{minipage}\\
\end{tabular}
\caption{Cosmic ray signals in the present model.
(a): positron fraction with experimental data\,\cite{Adriani:2008zr,Aguilar:2007yf,Barwick:1997ig}.
(b): positron and electron fluxes with experimental data\,\cite{Abdo:2009zk,Collaboration:2008aaa,Chang:2008zz,Torii:2008xu}.
The yellow zone shows a systematic error and the dashed line shows the background flux.
}
\label{fig:electron}
\end{figure}

In Fig.\,\ref{fig:electron}, we show the predicted positron fraction (left)
and the electron plus positron total flux (right). 
For the analysis on the propagation of the cosmic ray in the galaxy, 
we adopt the same set-up in Ref.\,\cite{Shirai:2009kh}
based on Refs.\,\cite{Hisano:2005ec,Ibarra:2008qg}, namely the MED diffusion model\,\cite{Delahaye:2007fr}
and the NFW dark matter profile\,\cite{Navarro:1996gj}. 
As for the electron and positron background, we borrowed  the estimation given in
Refs.\,\cite{Moskalenko:1997gh,Baltz:1998xv}, with a normalization
factor $k_{\rm bg} = 0.65$.  
In the analysis of the positron fraction, 
we have taken into account 
the solar modulation effect in the current solar cycle\,\cite{Baltz:1998xv},
which affects the fraction in $E\lesssim 10$\,GeV.
The figures show that the excesses observed in the PAMELA and FERMI experiments 
are nicely fitted by both the annihilation/decay scenario with either four electron or four
muon final states.

\subsubsection{Flaton decays into SSM particles}\label{sec:ssm}
In the above analysis, so far, we have considered that the dark matters mainly
annihilate/decay into pairs of R-axions which eventually decay into muon or electron pairs.
In the Nambu-Goldstone dark matter scenario, however, the interactions between
the flaton and the SSM fields are not arbitrary but are fixed for a given 
gauge mediation mechanism.
Thus, in the Nambu-Goldstone dark matter scenario, the SSM final states
of the dark matter annihilation/decay are definitely determined for a given gauge 
mediation model, which gives the model a distinctive prediction on the gamma-ray
spectrum.

The decay modes of the flaton into the SSM particles are given as follows.  
Since we are assuming the models with gauge mediation throughout the paper,
the interactions between the flaton and the SSM fields are obtained along with
the gauge mediation effects.
For example, the effective coupling between the flaton and
the gauginos is given by a Yukawa interaction;
\begin{eqnarray}
\label{eq:gaugino}
 {\cal L}_{\rm eff} 
  \simeq \frac{1}{2}\,
  \frac{\partial m_i}{\partial s}\,
s \,\l^{i}\l^{i} 
 + h.c.,
\end{eqnarray}
where $m_{i}$ denotes the gaugino mass and $i$ runs the SSM gauge
groups. 
For example, in a class of the so-called minimal gauge mediation\,\cite{Giudice:1998bp}, 
we obtain,
\begin{eqnarray}
   \frac{\partial m_i}{\partial s} = \frac{m_i}{f_R}.
\end{eqnarray}
In generic gauge mediation models,
it is expected
\begin{eqnarray}
   \frac{\partial m_i}{\partial s} = c_g \frac{m_i}{f_R},
\end{eqnarray}
with an order one coefficient $c_g$.%
\footnote{In a class of model based on the messenger sector\,\cite{Izawa:1997gs},
the coefficient $c_g$ can be vanishing for a special value of $f_R$ 
(see Fig.4 in Ref.\,\cite{Nomura:1997uu}). }

The coupling between the flaton and
the sfermions and Higgs bosons are also obtained in a similar way;
\begin{eqnarray}
\label{eq:scalar}
{\cal L}_{\rm eff} = \left.\frac{\partial m_{\tilde f}^{2}}{\partial s}\right|_{s=0} \times \,s\,\tilde f \tilde f,
\end{eqnarray}
and the model dependent coefficient ${\partial m_{\tilde f}^{2}}/{\partial s}$ is again given by,
\begin{eqnarray}
 \left.\frac{\partial m_{\tilde f}^{2}}{\partial s}\right|_{s=0} = \frac{m_{\tilde f}^2}{f_R},
\end{eqnarray}
for the minimal gauge mediation model.

With these interactions, the flaton decays into a pair of the SSM particles. 
For instance, the decay rate into a pair of the gluinos are given by
\begin{eqnarray}
\label{eq:gaugino2}
 \G_{s\to \tilde g\tilde g} \simeq \frac{c_g^2}{4\pi} \left( \frac{m_{\tilde g}}{m_{s}} \right)^{2}
 \frac{m_{s}^{3} }{f_{R}^{2}}\ .
\end{eqnarray}
Thus, when the gluino mass is close to that of the flaton, 
the branching ratio of the flaton into the gluino pairs can be sizable,
which gives a non-trivial contribution to the  gamma-ray spectrum.%
\footnote{
Notice that the branching ratios 
into the gauge boson pairs are highly suppressed\,\cite{Ibe:2006rc}.
}

In the rest of this section, we demonstrate how the SSM modes affect
the cosmic gamma-ray spectrum by concentrating on the effects of the 
sizable gluino branching fraction and take the gluino branching ratio,
\begin{eqnarray}
\label{eq:brag}
{\rm Br}_{\tilde{g}} = \frac{\G_{s\to \tilde g\tilde g}}{ \G_{s\to aa} }
= 8 c_g^2  \left( \frac{m_{\tilde g}}{m_{s}} \right)^{2}
\simeq 5.6\times 10^{-2}\times c_g^2\,
\left(\frac{m_{\tilde g}}{500\,{\rm GeV}} \right)^2
\left(\frac{6\,{\rm TeV}}{m_s}\right)^2
 \ ,
\end{eqnarray}
as a free parameter, which we assume is typically below 10\,\%.%
\footnote{
In the decaying dark matter scenario, $m_s$ in Eq.\,(\ref{eq:brag})
is replaced by $m_\phi$.}
The following analysis can be extended straightforwardly to more detailed analysis
on model by model basis for a given gauge mediation mechanism, although 
we do not peruse in this study.

\subsubsection{Gamma-ray and antiproton spectrum with a gluino mode}
The gamma-ray signals come from the final state radiation (FSR), inverse Compton scattering (ICS)
and the fragmentations of the final states of the decay/annihilation of the dark matter.
Especially, in the present model, the contribution from the
final states with gluinos has characteristic feature in high-energy region.
(As for the FSR and ICS contribution, see Refs.\,\cite{Bergstrom:2008ag,Mardon:2009gw} and \cite{Cirelli:2009vg,Ishiwata:2009dk}, respectively)

In this study, we concentrate on the gamma-ray signal from the fragmentation
of the gluino final states.
As a demonstration, we assume that the branching ratio to the gluinos 
of the dark matter decay/annihilation is 5\,\%.
Notice that the gamma-ray signal also depends on the decay mode of the gluino.
Here, we assume that the sfermions are heavier than the gauginos,
and consider two cases with
$m_{\tilde{\chi}^0_1}=100$ GeV and $450$ GeV, 
while the gluino mass is fixed to $m_{\tilde g}=500$\,GeV.
For simplicity, we further assume the case that the gluino decays into only light quark pair 
$q,\bar{q}$ and the neutralino $\tilde{\chi}^0_1$, 
and the neutralino $\tilde{\chi}^0_1$ decays into a gravitino and photon.
\begin{figure}[t!]
\begin{tabular}{lr}
\begin{minipage}{0.5\hsize}
\begin{center}
\epsfig{file=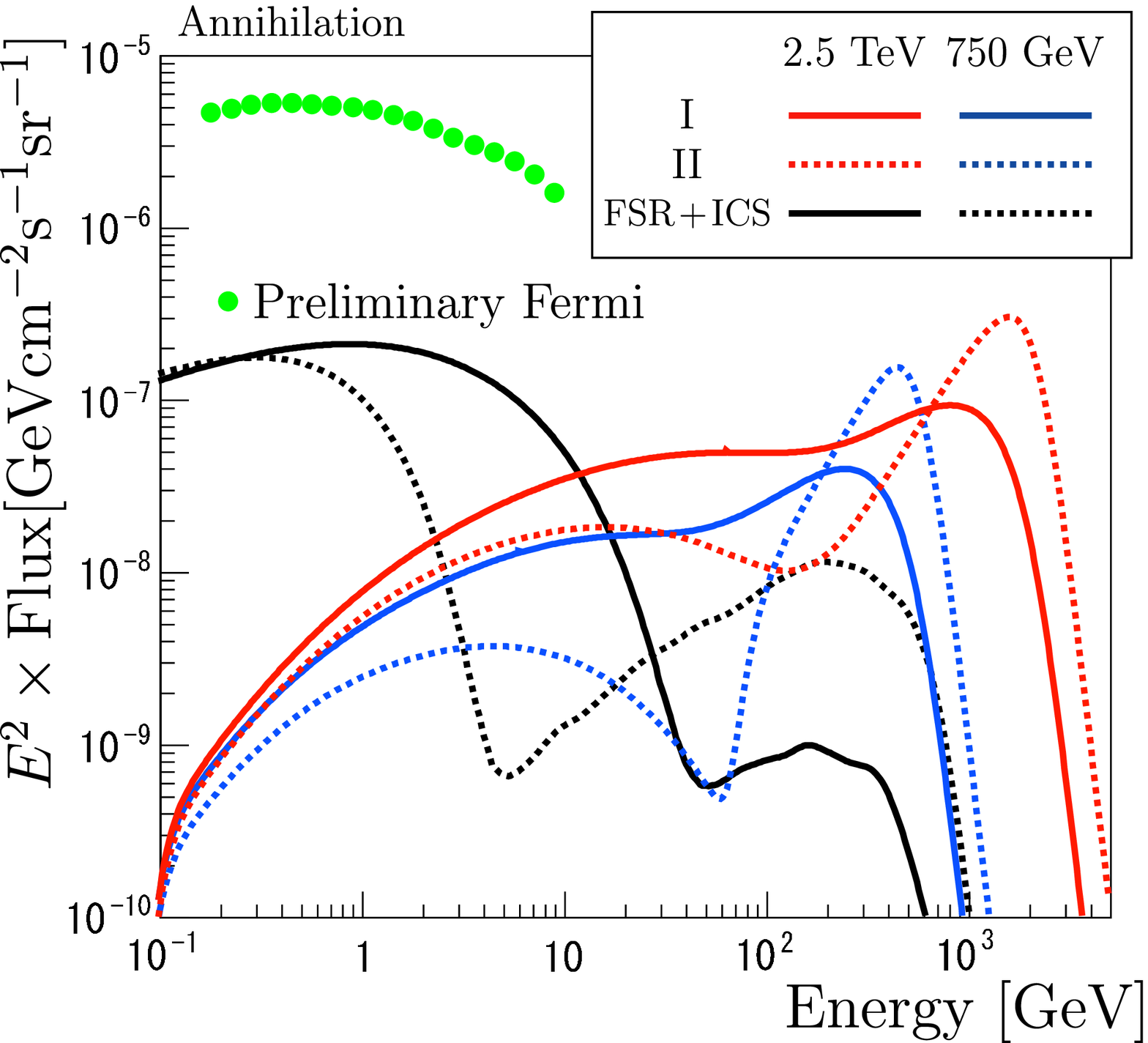 ,scale=.42,clip}\\
(a) Annihilation
\end{center}
\end{minipage}
\begin{minipage}{0.5\hsize}
\begin{center}
\epsfig{file=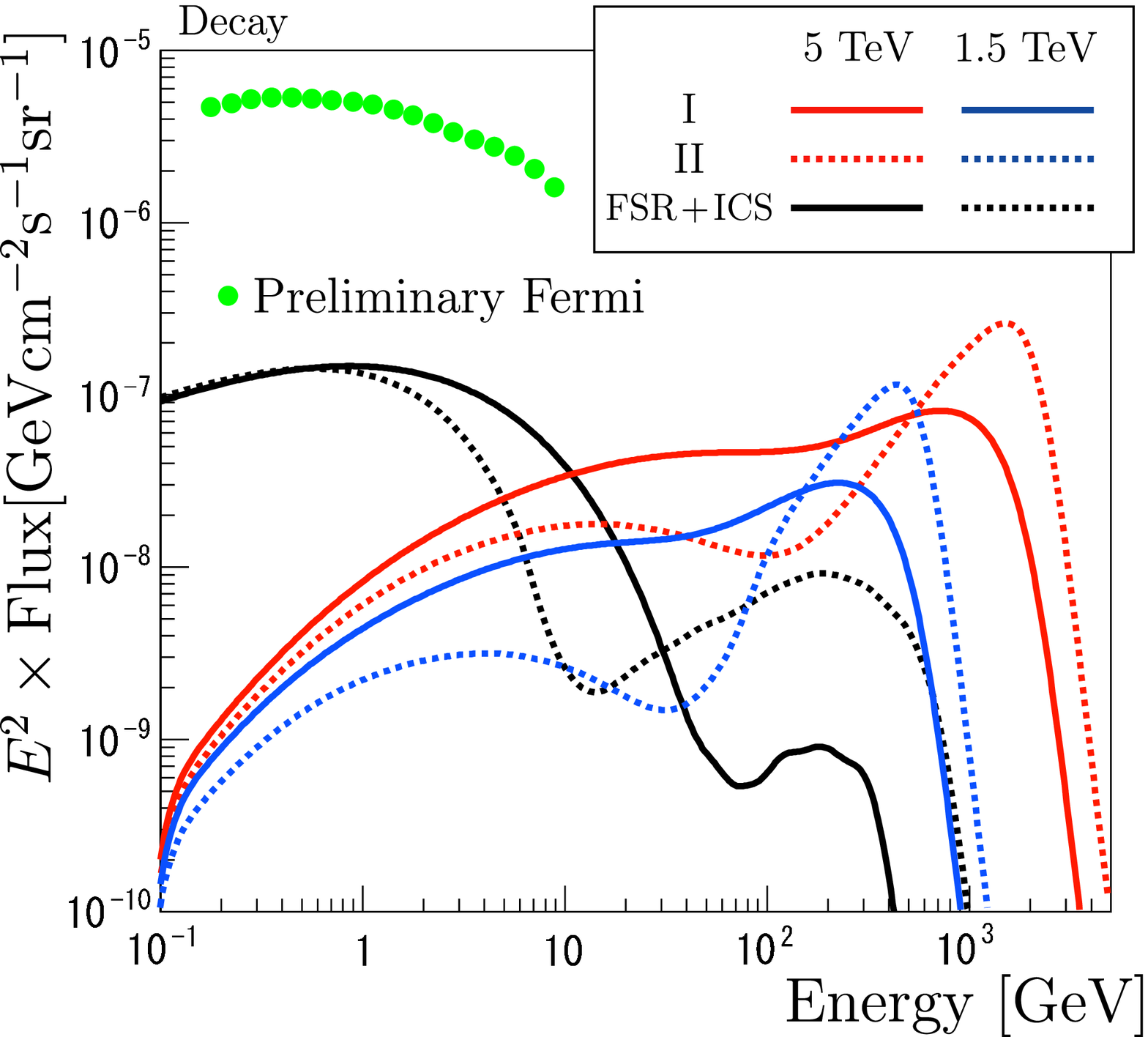 ,scale=.42,clip}\\
(b) Decay
\end{center}
\end{minipage}\\
\end{tabular}
\caption{Gamma-ray signals  in the present model with experimental data ($0 ^\circ\le \ell \le 360 ^\circ, ~ 10^\circ\le|b|\le 20^\circ$)\,\cite{Porter:2009sg}.
I and II represent the SSM spectrum where $m_{\tilde{\chi}^0_1} = 100$ GeV and 450 GeV, respectively.
FSR+ICS represents the contributions of the FSR and ICS of the electron and positron cosmic ray.
}
\label{fig:gamma}
\end{figure}
In Fig.~\ref{fig:gamma}, the gamma-ray fluxes are shown.
To estimate the fluxes, we have used the NFW profile 
and averaged the halo signal over the region $0 ^\circ\le \ell \le 360 ^\circ, ~ 10^\circ\le|b|\le 20^\circ$
For the annihilation case, we only include the halo component.%
\footnote{
As for extra-galactic gamma-ray, see e.g., Ref.\,\cite{Kawasaki:2009nr}
}
In Fig.~\ref{fig:gamma}, we also show the FSR and ICS contributions from the electron and positron which come from
the DM dominant annihilation/decay mode.
We estimate the ICS component with the method discussed in Ref.\,\cite{Ishiwata:2009dk},
using data of interstellar radiation field provided by the GALPROP collaboration\,\cite{GALPROP}, which is based on Ref.\,\cite{Porter:2005qx}.

In both cases, the DM signals are consistent with the current experiment data,
and anomalous behavior of the gamma ray is expected around the DM mass for the annihilation cases or half for decay.
This behavior comes from the hard component from the neutralino $\tilde{\chi}^0_1$ decay.

\begin{figure}[t!]
\begin{tabular}{lr}
\begin{minipage}{0.5\hsize}
\begin{center}
\epsfig{file=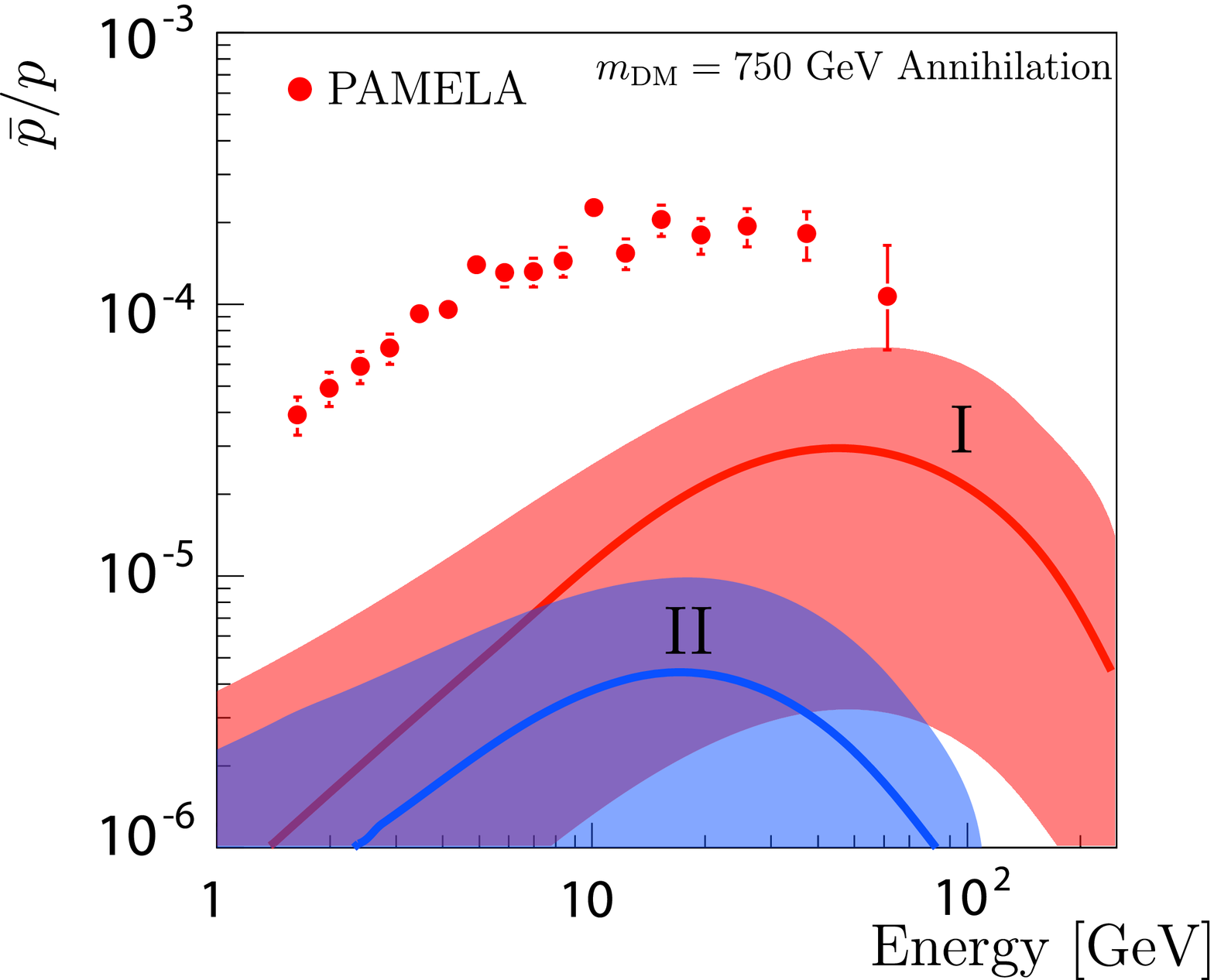 ,scale=.42,clip}\\
(a) Annihilation $m_{\rm DM} = 750$ GeV.
\end{center}
\end{minipage}
\begin{minipage}{0.5\hsize}
\begin{center}
\epsfig{file=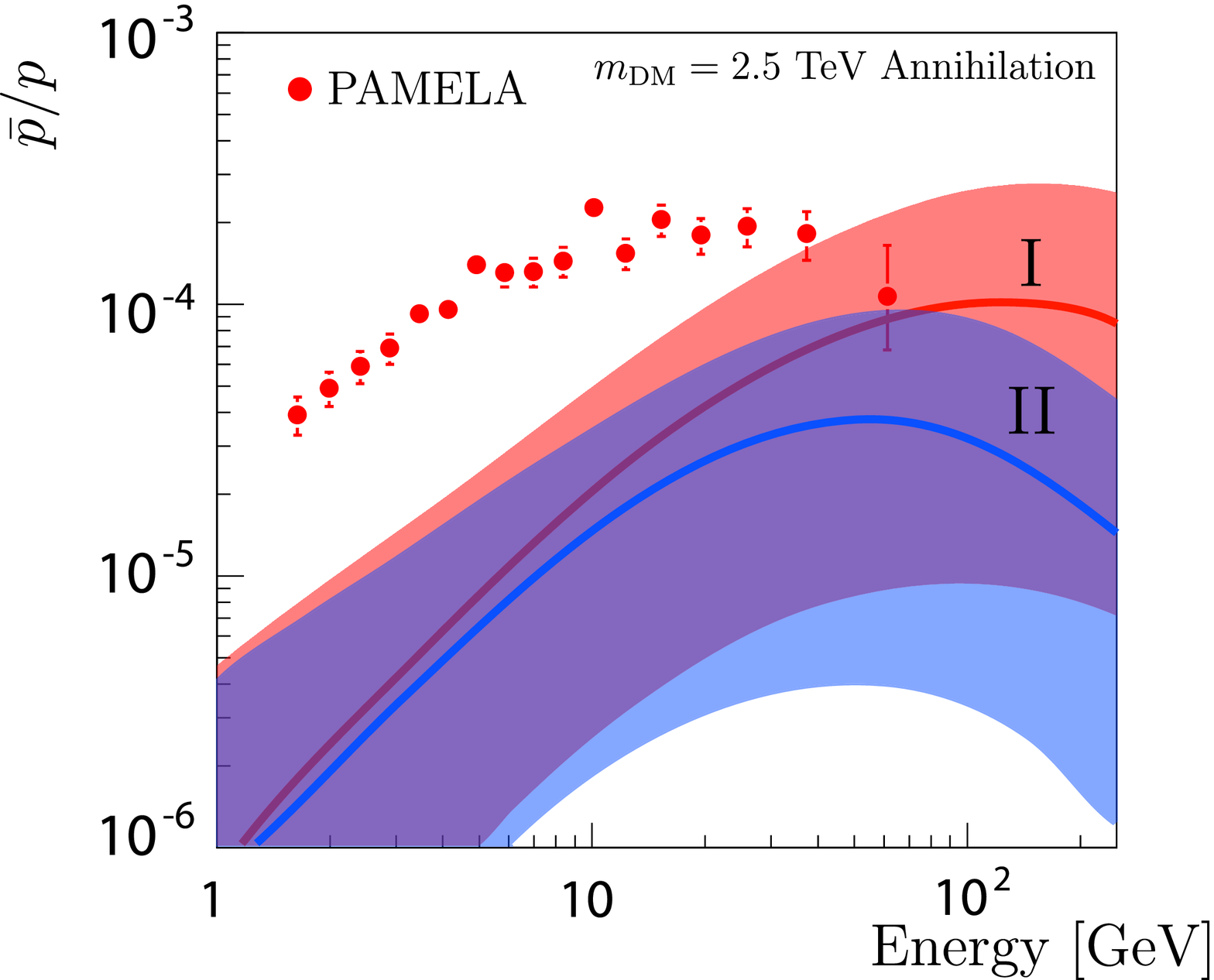 ,scale=.42,clip}\\
(b)  Annihilation $m_{\rm DM} = 750$ GeV.
\end{center}
\end{minipage}\\
\begin{minipage}{0.5\hsize}
\begin{center}
\epsfig{file=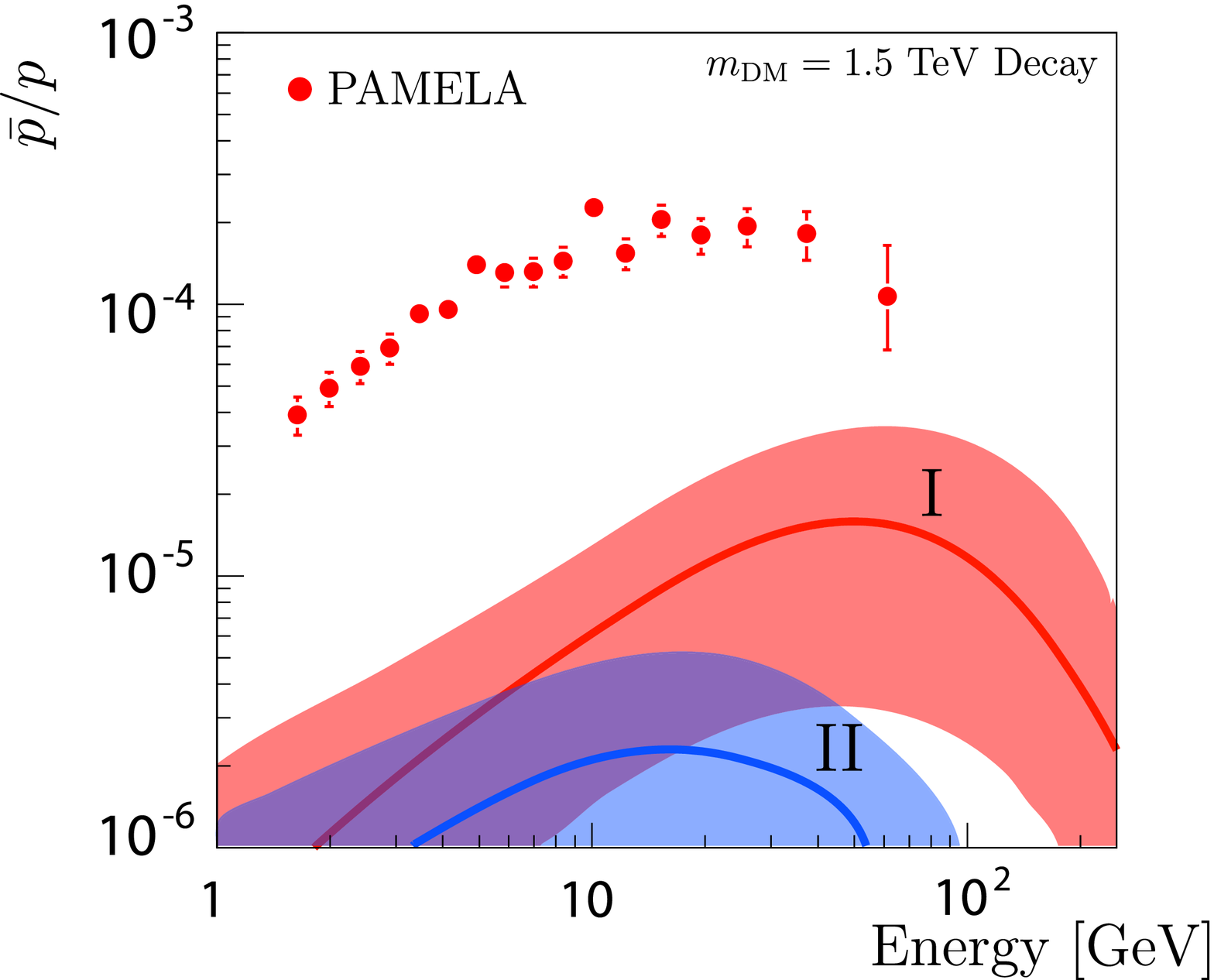 ,scale=.42,clip}\\
(c)  Decay $m_{\rm DM} = 1500$ GeV.
\end{center}
\end{minipage}
\begin{minipage}{0.5\hsize}
\begin{center}
\epsfig{file=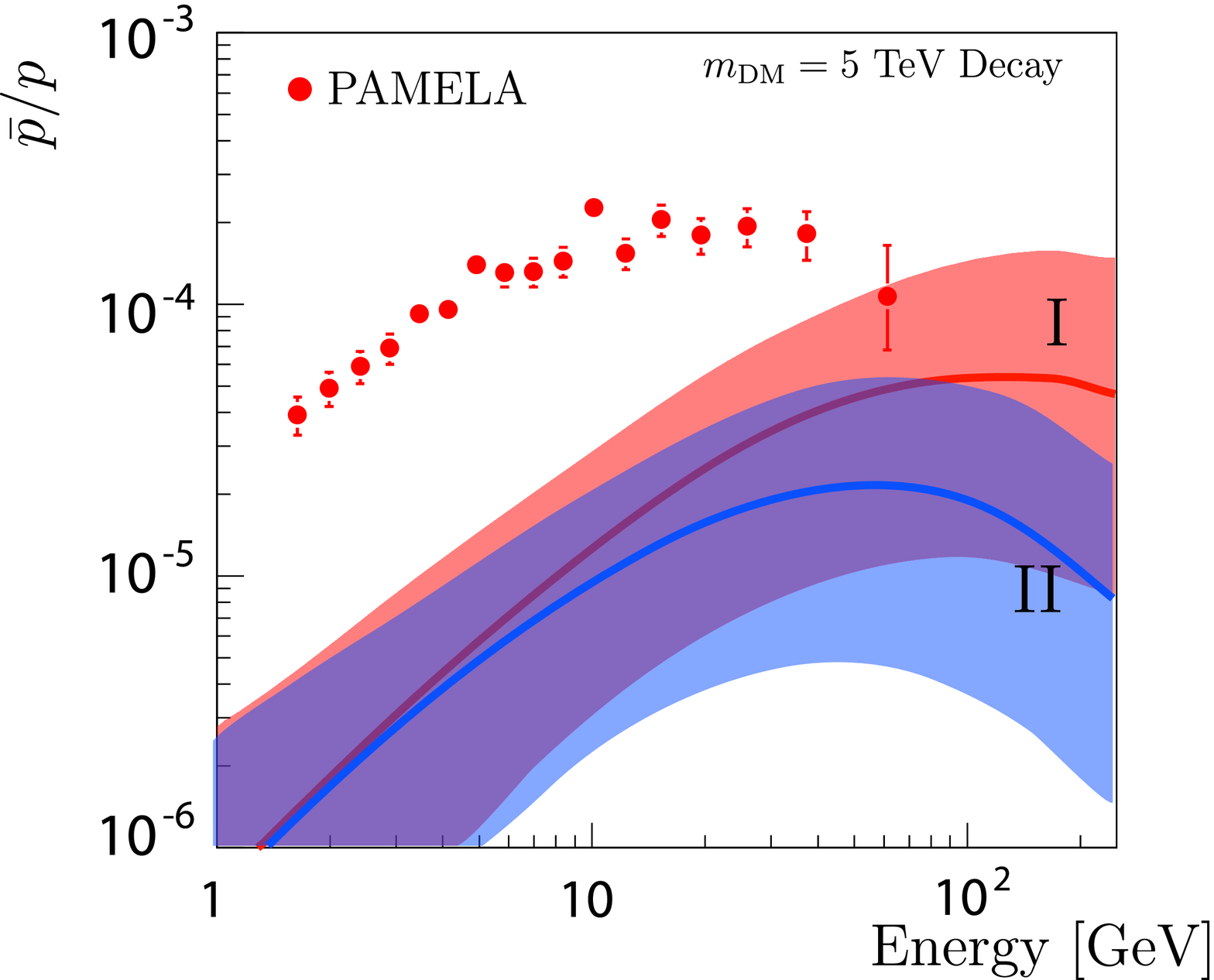 ,scale=.42,clip}\\
(d)  Decay $m_{\rm DM} = 5000$ GeV.
\end{center}
\end{minipage}\\

\end{tabular}
\caption{Antiproton signals in the present model with experimental data\,\cite{Adriani:2008zq}.
I and II represent the SSM spectrum where $m_{\tilde{\chi}^0_1} = 100$ GeV and 450 GeV, respectively.
}
\label{fig:anti}
\end{figure}

The gluino final states also  raise the antiproton signal, which is severely  constrained 
by the PAMELA experiment\,\cite{Adriani:2008zq}.
In Fig.~\ref{fig:anti}, we show the ratio of antiproton which comes from signal and the background proton.
In this analysis, we used the program PYTHIA\,\cite{Sjostrand:2006za}
to obtain the fragmentation function of the final states into antiprotons.
The analysis of the propagation of the antiproton is again based on that
used in Ref.\,\cite{Shirai:2009kh}.
The figure shows that the antiproton flux strongly depends on the diffusion models,
and the shaded region corresponds to the dependence on the diffusion model,
with upper side of the region corresponds to the diffusion model MAX, under side MIN
and the line MED in Ref. \,\cite{Delahaye:2007fr}.
The proton background is  taken from Ref.\,\cite{Chen:2009gz}.
The larger antiproton fluxes in the case of  ``I" reflect the higher energy fractions of the
quark pairs in the decay of the gluinos.
We see the contradiction between the experiments and the signals in some diffusion models
even for the 5\,\% branching ratio into the gluino final states.

As a result, we see that the antiproton flux provides very strong constraint on the model.
For example, in the minimal gauge mediation, the branching fraction 
into the gluino final states are solely determined by the masses
of the dark matter and the gluino, {\it i.e.} $c_g =1$ in Eq.\,(\ref{eq:brag}).
Thus, the strict constraint on the branching ratio leads to a strict constraint 
on the mass of the gluino which is an important parameter
for the SUSY search at the coming LHC experiments.
Therefore, the Nambu-Goldstone dark matter scenario can be
investigated with the interplay between the cosmic-ray experiments as well 
as the direct SUSY search at the LHC experiment.



\section*{Acknowledgements}
The work of M.~I. was supported by the U.S. Department of Energy under
contract number DE-AC02-76SF00515.  
The work of H.M. and T.T.Y. was supported in
part by World Premier International Research Center Initiative (WPI
Initiative), MEXT, Japan. The work of H.M. was also supported in part
by the U.S. DOE under Contract DE-AC03-76SF00098, and in part by the
NSF under grant PHY-04-57315.
The work of SS is supported in part by JSPS Research Fellowships for Young
Scientists.


\begin{thebibliography}{99}  


\bibitem{Adriani:2008zr}
  O.~Adriani {\it et al.}  [PAMELA Collaboration],
  Nature {\bf 458}, 607 (2009)
  [arXiv:0810.4995 [astro-ph]].
\bibitem{Chang:2008zz}
  J.~Chang {\it et al.},
  Nature {\bf 456}, 362 (2008).
\bibitem{Torii:2008xu}
  S.~Torii {\it et al.},
  arXiv:0809.0760 [astro-ph].
 
\bibitem{Abdo:2009zk}
  A.~A.~Abdo {\it et al.}  [The Fermi LAT Collaboration],
  arXiv:0905.0025 [astro-ph.HE].
\bibitem{Bergstrom:2009ib}
For a review, see, {\it e.g.},   L.~Bergstrom,
  arXiv:0903.4849 [hep-ph] and references therein.
 \bibitem{Murayama:2009}
H.~Murayama, talk at 2009 APS  April Meeting, May 3, 2009, Denver, Colorado.   
\bibitem{Bergstrom:2009fa}
  L.~Bergstrom, J.~Edsjo and G.~Zaharijas,
  arXiv:0905.0333 [astro-ph.HE].
\bibitem{Meade:2009iu}
  P.~Meade, M.~Papucci, A.~Strumia and T.~Volansky,
  arXiv:0905.0480 [hep-ph].
\bibitem{ArkaniHamed:2008qn}
  N.~Arkani-Hamed, D.~P.~Finkbeiner, T.~R.~Slatyer and N.~Weiner,
  Phys.\ Rev.\  D {\bf 79}, 015014 (2009)
  [arXiv:0810.0713 [hep-ph]].
\bibitem{Nomura:2008ru}
  Y.~Nomura and J.~Thaler,
  arXiv:0810.5397 [hep-ph].

\bibitem{Murayama:2007ek}
See, {\it e.g.},  H.~Murayama,
  arXiv:0704.2276 [hep-ph].
  
\bibitem{Shirai:2009fq}
  S.~Shirai, F.~Takahashi and T.~T.~Yanagida,
  arXiv:0905.0388 [hep-ph].
\bibitem{Chen:2009mj}
  C.~H.~Chen, C.~Q.~Geng and D.~V.~Zhuridov,
  arXiv:0905.0652 [hep-ph].

\bibitem{Ibe:2009dx}
  M.~Ibe, Y.~Nakayama, H.~Murayama and T.~T.~Yanagida,
  JHEP {\bf 0904}, 087 (2009)
  [arXiv:0902.2914 [hep-ph]].
  
\bibitem{Giudice:1998bp}
For a review, see, {\it e.g.},  G.~F.~Giudice and R.~Rattazzi,
  Phys.\ Rept.\  {\bf 322}, 419 (1999)
  [arXiv:hep-ph/9801271],
  and references therein.

  
\bibitem{Izawa:1996pk}
  K.~I.~Izawa and T.~Yanagida,
  Prog.\ Theor.\ Phys.\  {\bf 95}, 829 (1996)
  [arXiv:hep-th/9602180];
  K.~A.~Intriligator and S.~D.~Thomas,
  Nucl.\ Phys.\  B {\bf 473}, 121 (1996)
  [arXiv:hep-th/9603158].
  
\bibitem{Bagger:1994hh}
  J.~Bagger, E.~Poppitz and L.~Randall,
  Nucl.\ Phys.\  B {\bf 426}, 3 (1994)
  [arXiv:hep-ph/9405345].


\bibitem{Goh:2008xz}
  H.~S.~Goh and M.~Ibe,
  arXiv:0810.5773 [hep-ph].
  
\bibitem{Bergsma:1985qz}
  F.~Bergsma {\it et al.}  [CHARM Collaboration],
  Phys.\ Lett.\  B {\bf 157}, 458 (1985).
  
\bibitem{Kim:1986ax}
  J.~E.~Kim,
  Phys.\ Rept.\  {\bf 150}, 1 (1987).


  \bibitem{Dine:2006xt}
  M.~Dine and J.~Mason,
  Phys.\ Rev.\  D {\bf 77}, 016005 (2008)
  [arXiv:hep-ph/0611312].


\bibitem{Griest:1990kh}
  K.~Griest and D.~Seckel,
  Phys.\ Rev.\  D {\bf 43}, 3191 (1991).
\bibitem{Gondolo:1990dk}
  P.~Gondolo and G.~Gelmini,
  Nucl.\ Phys.\  B {\bf 360}, 145 (1991).



\bibitem{Komatsu:2008hk}
  E.~Komatsu {\it et al.}  [WMAP Collaboration],
  arXiv:0803.0547 [astro-ph].

\bibitem{Ibe:2008ye}
  M.~Ibe, H.~Murayama and T.~T.~Yanagida,
  Phys.\ Rev.\  D {\bf 79}, 095009 (2009)
  [arXiv:0812.0072 [hep-ph]].
\bibitem{CyrRacine:2009yn}
  F.~Y.~Cyr-Racine, S.~Profumo and K.~Sigurdson,
  arXiv:0904.3933 [astro-ph.CO].


\bibitem{Aguilar:2007yf}
  M.~Aguilar {\it et al.}  [AMS-01 Collaboration],
  Phys.\ Lett.\  B {\bf 646}, 145 (2007)
  [arXiv:astro-ph/0703154].

\bibitem{Barwick:1997ig}
  S.~W.~Barwick {\it et al.}  [HEAT Collaboration],
  Astrophys.\ J.\  {\bf 482}, L191 (1997)
  [arXiv:astro-ph/9703192].


  \bibitem{Collaboration:2008aaa}
  F.~Aharonian {\it et al.}  [H.E.S.S. Collaboration],
  Phys.\ Rev.\ Lett.\  {\bf 101}, 261104 (2008)
  [arXiv:0811.3894 [astro-ph]];
  arXiv:0905.0105 [astro-ph.HE].


\bibitem{Shirai:2009kh}
  S.~Shirai, F.~Takahashi and T.~T.~Yanagida,
  arXiv:0902.4770 [hep-ph].

\bibitem{Hisano:2005ec}
  J.~Hisano, S.~Matsumoto, O.~Saito and M.~Senami,
  Phys.\ Rev.\  D {\bf 73}, 055004 (2006)
  [arXiv:hep-ph/0511118].
  
\bibitem{Ibarra:2008qg}
  A.~Ibarra and D.~Tran,
  JCAP {\bf 0807}, 002 (2008)
  [arXiv:0804.4596 [astro-ph]].
  

\bibitem{Delahaye:2007fr}
  T.~Delahaye, R.~Lineros, F.~Donato, N.~Fornengo and P.~Salati,
  Phys.\ Rev.\  D {\bf 77}, 063527 (2008)
  [arXiv:0712.2312 [astro-ph]].
  
\bibitem{Navarro:1996gj}
  J.~F.~Navarro, C.~S.~Frenk and S.~D.~M.~White,
  Astrophys.\ J.\  {\bf 490}, 493 (1997)
  [arXiv:astro-ph/9611107].


\bibitem{Moskalenko:1997gh}
  I.~V.~Moskalenko and A.~W.~Strong,
  Astrophys.\ J.\  {\bf 493} (1998) 694
  [arXiv:astro-ph/9710124].


\bibitem{Baltz:1998xv}
  E.~A.~Baltz and J.~Edsjo,
  Phys.\ Rev.\  D {\bf 59} (1999) 023511
  [arXiv:astro-ph/9808243].
  
  
\bibitem{Izawa:1997gs}
  K.~I.~Izawa, Y.~Nomura, K.~Tobe and T.~Yanagida,
  Phys.\ Rev.\  D {\bf 56}, 2886 (1997)
  [arXiv:hep-ph/9705228].
\bibitem{Nomura:1997uu}
  Y.~Nomura and K.~Tobe,
  Phys.\ Rev.\  D {\bf 58}, 055002 (1998)
  [arXiv:hep-ph/9708377].
  





\bibitem{Ibe:2006rc}
  M.~Ibe and R.~Kitano,
  Phys.\ Rev.\  D {\bf 75}, 055003 (2007)
  [arXiv:hep-ph/0611111].

\bibitem{Bergstrom:2008ag}
  L.~Bergstrom, G.~Bertone, T.~Bringmann, J.~Edsjo and M.~Taoso,
  arXiv:0812.3895 [astro-ph].

\bibitem{Mardon:2009gw}
  J.~Mardon, Y.~Nomura and J.~Thaler,
  arXiv:0905.3749 [hep-ph].

\bibitem{Cirelli:2009vg}
  M.~Cirelli and P.~Panci,
  arXiv:0904.3830 [astro-ph.CO].

\bibitem{Ishiwata:2009dk}
  K.~Ishiwata, S.~Matsumoto and T.~Moroi,
  arXiv:0905.4593 [astro-ph.CO].

  

\bibitem{Kawasaki:2009nr}
  M.~Kawasaki, K.~Kohri and K.~Nakayama,
  arXiv:0904.3626 [astro-ph.CO].

\bibitem{Porter:2009sg}
  T.~A.~Porter and f.~t.~F.~Collaboration,
  arXiv:0907.0294 [astro-ph.HE].

\bibitem{GALPROP}
GALPROP Homepage, http://galprop.stanford.edu/.
\bibitem{Porter:2005qx}
  T.~A.~Porter and A.~W.~Strong,
  arXiv:astro-ph/0507119.


\bibitem{Adriani:2008zq}
  O.~Adriani {\it et al.},
  Phys.\ Rev.\ Lett.\  {\bf 102}, 051101 (2009)
  [arXiv:0810.4994 [astro-ph]].


\bibitem{Sjostrand:2006za}
  T.~Sjostrand, S.~Mrenna and P.~Skands,
  JHEP {\bf 0605}, 026 (2006)
  [arXiv:hep-ph/0603175].



\bibitem{Chen:2009gz}
  C.~R.~Chen, M.~M.~Nojiri, S.~C.~Park, J.~Shu and M.~Takeuchi,
  arXiv:0903.1971 [hep-ph].


\end{thebibliography}
\end{document}